\documentclass[sts]{imsart}

\RequirePackage[colorlinks,citecolor=blue,urlcolor=blue]{hyperref}

\startlocaldefs
\usepackage{amsfonts}
\usepackage{amsmath}
\usepackage{amssymb}
\usepackage{amsthm}
\usepackage{subcaption}
\usepackage{suffix}
\usepackage{color}
\usepackage{bigstrut}
\usepackage{graphicx}
\usepackage{wrapfig}
\usepackage{epstopdf}
\usepackage{multirow}
\usepackage{nicefrac}
\usepackage[normalem]{ulem}
\usepackage[inline]{enumitem}
\setlist*[enumerate]{label=(\roman*)}

\usepackage{natbib}

\graphicspath{{./figs/}}

\newcommand{\y}{{\bf y}}
\newcommand{\D}{{\bf D}}
\newcommand{\E}{\mathbb E}

\newcommand{\I}{{\bf I}}

\newcommand{\M}{{\bf M}}
\newcommand{\Nor}{{\cal N}}  

\newcommand{\U}{{\bf U}}

\newcommand{\W}{{\bf W}}
\newcommand{\X}{{\bf X}}
\newcommand{\Y}{{\bf Y}}
\newcommand{\Z}{{\bf Z}}

\newcommand{\balpha}{\boldsymbol{\alpha}}

\newcommand{\blambda}{\boldsymbol{\lambda}}

\newcommand{\btheta}{\boldsymbol{\theta}}

\newcommand{\bSigma}{\boldsymbol{\Sigma}}

\newcommand{\bgamma}{\boldsymbol{\gamma}}

\newcommand{\bepsilon}{\boldsymbol{\epsilon}}

\newcommand{\1}{{\bf 1}}
\newcommand{\0}{{\bf 0}}

\newcommand{\ben}{\begin{enumerate}}
\newcommand{\een}{\end{enumerate}}
\newcommand{\bit}{\begin{itemize}}
\newcommand{\eit}{\end{itemize}}
\newcommand{\beq}{\begin{equation}}
\newcommand{\eeq}{\end{equation}}
\newcommand{\bde}{\begin{description}}
\newcommand{\ede}{\end{description}}

\newcommand{\half}{\frac{1}{2}}

\newcommand{\argmin}{\operatornamewithlimits{argmin}}
\newcommand{\argmax}{\operatornamewithlimits{argmax}}

\newcommand{\abs}[1]{\lvert#1\rvert}
\newcommand{\norm}[1]{\lVert#1\rVert}
\newcommand{\vectornorm}[1]{\left|\left|#1\right|\right|}

\newcommand{\dd}[2]{\frac{\partial #1}{\partial #2}}

\newcommand{\id}{\stackrel{\mathrm{ind}}{\sim}}

\newcommand{\defeq}{\operatorname{:=}}

\newcommand{\estHs}{\ensuremath{{\hat{\theta}}_{HS}}{}}
\newcommand{\estJs}{\ensuremath{{\hat{\btheta}}_{JS}}{}}

\newtheorem{theorem}{Theorem}

\newtheorem{definition}[theorem]{Definition}

\newtheorem{lemma}[theorem]{Lemma}

\newtheorem{remark}[theorem]{Remark}

\endlocaldefs

\begin{document}

\begin{frontmatter}

\title{Lasso Meets Horseshoe: A Survey}
\runtitle{Lasso Meets Horseshoe: A Survey}


\author{\fnms{Anindya} \snm{Bhadra}\ead[label=e1]{bhadra@purdue.edu}}
\address{250 N. University St., West Lafayette, IN 47907. \printead{e1}}
\vspace{-0.08in}
\affiliation{Purdue University}
\author{\fnms{Jyotishka} \snm{Datta}\corref{}\ead[label=e2]{jd033@uark.edu}}
\address{1 University of Arkansas, Fayetteville, AR 72704. \printead{e2}}
\affiliation{University of Arkansas}
\author{\fnms{Nicholas G.} \snm{Polson}\ead[label=e3]{ngp@chicagobooth.edu}}
\address{5807 S. Woodlawn Ave., Chicago, IL 60637. \printead{e3}}
\affiliation{University of Chicago  \\   Booth School of Business}
\author{\fnms{Brandon} \snm{Willard}\ead[label=e4]{bwillard@uchicago.edu}}
\address{5807 S. Woodlawn Ave., Chicago, IL 60637. \printead{e4}}
\affiliation{University of Chicago  \\   Booth School of Business}

\runauthor{Bhadra, Datta, Polson and Willard}

\begin{abstract}
The goal of this paper is to contrast and survey the major advances in two of the most commonly used high-dimensional techniques, namely, the Lasso and horseshoe regularization. Lasso is a gold standard for predictor selection while horseshoe is a state-of-the-art Bayesian estimator for sparse signals. Lasso is fast and scalable and uses convex optimization whilst the horseshoe is non-convex.  Our novel perspective focuses on three aspects: 
\begin{enumerate*}
  \item theoretical optimality in high-dimensional inference for the Gaussian sparse model and beyond, 
  \item efficiency and scalability of computation and 
  \item methodological development and performance. 
\end{enumerate*} 
\end{abstract}



\end{frontmatter}

\section{Introduction}

High-dimensional predictor selection and sparse signal recovery are routine statistical and machine learning practices. There is a vast and growing literature focusing on computational aspects of large scale inference problems. Whilst this area is too large to review here, we revisit two popular sparse parameter estimation techniques, the Lasso \citep{tibshirani96} and the horseshoe estimator \citep{carvalho2010horseshoe}. Specifically, we focus on three areas: performance in high-dimensional data, theoretical optimality and computational efficiency. 

Sparsity relies on the property of a few large signals among many (nearly) zero noisy observations. A common goal in high-dimensional inference is to recover the low-dimensional signals observed in noisy observations. This problem encompasses four related areas: 

\begin{enumerate}
  \item Estimation of the underlying sparse parameter vector. 
  \item Multiple testing where the \# tests is much larger than the sample size, $n$. 
  \item Regression subset selection where \# of covariates $p$ is far larger than $n$. 
  \item Out-of-sample prediction. 
\end{enumerate}

There are a rich variety of methodologies for high-dimensional regularization which implicitly or explicitly penalize model dimensionality. Lasso (Least Absolute Shrinkage and Selection Operator) produces a sparse estimate by constraining the $\ell_1$ norm of the parameter vector. Lasso's widespread popularity is due to a multitude of factors, in particular due to the computational efficiency of the least angle regression (LARS) \citep{efron_least_2004} or the simple coordinate descent approaches of \citet{friedman_pathwise_2007}, and its ability to produce a sparse solution, with optimality (oracle) properties for both estimation and variable selection \citep[\textit{vide}][]{buhlmann2011statistics, james2013introduction, hastie2015statistical}. Table~\ref{tab:lasso:ext}, adapted from \citet{tibshirani2014praise},  gives a list of popular regularization methods based on Lasso.
\par 

\begin{table}[ht!]
  \centering
  \caption{Lasso regularization methods}
  \footnotesize{
    \begin{tabular}{|c|c|}
    \hline
    Method  & Authors  \bigstrut\\
    \hline
    Adaptive Lasso & \citet{zou2006adaptive} \bigstrut[t]\\
    Compressive sensing  & \citet{donoho2006compressed,candes2008restricted} \\
    Dantzig selector  & \citet{candes2007dantzig} \\
    Elastic net & \citet{zou2005regularization} \\
    Fused Lasso & \citet{tibshirani_sparsity_2005} \\
    Generalized Lasso & \citet{tibshirani2011solution} \\
    Graphical Lasso & \citet{friedman2008sparse} \\
    Grouped Lasso & \citet{yuan2006model} \\
    Hierarchical interaction models & \citet{bien_lasso_2013} \\
    Matrix completion & \citet{candes2010power,mazumder2010spectral} \\
    Multivariate methods & \citet{jolliffe2003modified,witten2009penalized} \\
    Near-isotonic regression & \citet{tibshirani2011nearly} \\
    Square Root Lasso  & \citet{belloni2011square} \\
    Scaled Lasso & \citet{sun2012scaled} \\
    Minimum concave penalty & \citet{zhang2010nearly} \\
    SparseNet & \citet{mazumder2012} \bigstrut[b]\\
    \hline
    \end{tabular}%
    }
  \label{tab:lasso:ext}%
\end{table}%

Bayes procedures, on the other hand, can be classified into two categories: two-groups model or spike-and-slab priors \citep{johnstone2004needles,efron2008microarrays,efron2010large,bogdan2011asymptotic,castillo2012needles} and global-local shrinkage priors \citep{carvalho2009handling,carvalho2010horseshoe,griffin2005alternative,armagan2011generalized,armagan2013generalized,polson2010shrink}, with the horseshoe prior \citep{carvalho2010horseshoe} being one of the most popular methods. The first class, spike-and-slab prior, places a discrete mixture of a point mass at zero (the spike) and an absolutely continuous density (the slab) on each parameter. The second entails placing absolutely continuous shrinkage priors on the entire parameter vector that selectively shrinks the small signals. Table \ref{tab:one-gps} provides a sampling of a few continuous shrinkage priors popular in the literature. Both these approaches have their own advantages and caveats, which we discuss in turn. A key duality is that the point estimate from a regularization approach can be interpreted as Bayesian mode of the posterior distribution under an appropriate shrinkage prior. 

\begin{table}[htbp]
  \centering
  \caption{A catalog of global-local shrinkage priors}
  \footnotesize{
    \begin{tabular}{|c|c|}
    \hline
    Global-local shrinkage prior  & Authors  \bigstrut\\
    \hline
    Normal Exponential Gamma & \citet{griffin2005alternative} \bigstrut[t]\\
    Horseshoe & \citet{carvalho2010horseshoe, carvalho2009handling} \\
    Hypergeometric Inverted Beta & \citet{polson2010large} \\
    Generalized Double Pareto & \citet{armagan2011generalized} \\
    Generalized Beta  & \citet{armagan2013generalized} \\
    Dirichlet--Laplace & \citet{bhattacharya2014dirichlet} \\
    Horseshoe+  & \citet{bhadra2015horseshoe+} \\
    Horseshoe-like & \citet{bhadra2017horseshoe} \\
    Spike-and-Slab Lasso & \citet{rovckova2016spike} \\
    R2-D2 & \citet{zhang2016high} \bigstrut[b]\\
    Inverse-Gamma-Gamma & \citet{bai2017inverse} \bigstrut[b]\\
    \hline
    \end{tabular}%
    }
  \label{tab:one-gps}%
\end{table}%

Both Lasso and horseshoe procedures come with strong theoretical guarantees for estimation, prediction and variable selection. Both procedures possess asymptotic oracle properties, i.e. identify the true non-zero coefficients as well as achieve the optimal estimation rate. The behavior of the Lasso estimator in terms of the risk properties has been studied in depth and has resulted in many methods aiming to improve certain features (see Table \ref{tab:lasso:ext}). On the other hand, horseshoe and other global-local priors have been shown to achieve optimality in variable selection, estimation and prediction, that we review in \S \ref{sec:stat-prop}, although theoretical studies of the continuous shrinkage priors is still an active area. 


The rest of the paper is organized as follows: Section \ref{sec:2} provides historical background for the normal means (a.k.a the Gaussian compound decision problem) and the sparse regression problems. Section \ref{sec:3} provides the link between regularization and optimization perspectives viewed through a probabilistic Bayesian lens. Section \ref{sec:stat-prop} compares and contrasts the statistical risk properties of Lasso and the horseshoe prior. Sections \ref{sec:5} and \ref{sec:horse-comp} discuss the issues of hyper-parameter selection and computational strategies. Section \ref{sec:simulation} provides two simulation experiments comparing the horseshoe prior with penalized regression methods for linear model and logistic regression with varying degree of dependence between predictors. We discuss applications of Lasso and the horseshoe in Section \ref{sec:app-ext} and provide directions for future work in Section \ref{sec:9}.

\section{Sparse Normal Means, Regression and Variable Selection}\label{sec:2}

\subsection{Sparse Normal Means} Suppose that we observe data from the probability model $(y_i \mid \theta_i)  \id \Nor (\theta_i,1)$ for $i = 1, \ldots, n$. Our primary inferential goal is to estimate the vector of normal means $ \btheta = ( \theta_1, \ldots , \theta_n )$ and a secondary goal is to simultaneously test if $\theta_i$'s are coming from a null distribution. We are interested in the sparse paradigm where a large proportion of the parameter vector contains zeros.  The `nearly black' \citep{donoho1992maximum} regime occurs when the parameter vector $\theta$ lies in the set $\ell_0[ p_n] \equiv \{ \theta : \# ( \theta_i \neq 0 ) \leq p_n \} $ with the upper bound on the number of non-zero parameter values $ p_n = o(n) $ as $ n \to \infty$.

A natural Bayesian solution for inference under sparsity is the two-groups model that puts a non-zero probability spike at zero and a suitable prior on the non-zero $\theta_i$'s (\textit{vide} Appendix \ref{sec:2gp}). The inference is then based on the posterior probabilities of non-zero $\theta_i$'s based on the discrete mixture model. The two-groups model possesses a number of frequentist and Bayesian optimality properties. \cite{johnstone2004needles} showed that a thresholding-based estimator for $\btheta$ under the two-groups model with an empirical Bayes estimate for the sparsity proportion attains the minimax rate in $\ell_q$ norm for $q \in (0,2]$ for $\btheta$ that are either nearly black or belong to an $\ell_p$ ball of `small' radius. \cite{castillo2012needles} treated a full Bayes version of the problem and again found an estimate that is minimax in $\ell_q$ norm for mean vectors that are either nearly black or have bounded weak $\ell_p$ norm for $p \in (0,2]$. 


\subsection{Sparse Linear Regression} 

A related inferential problem is high-dimensional linear regression with sparsity constraints on the parameter vector $\btheta$. We are interested in the linear regression model:

\[
\Y = \X \btheta + \bepsilon,
\]

where $\X = [\X_1, \cdots, \X_p]$ is a $n \times p$ matrix of predictors and $\bepsilon \sim \Nor(\0, \sigma^2\I_n)$. 

Our focus is on the sparse solution where $p \gg n$ and most of $\theta_i$'s are zero. Similar to the sparse normal means problem, our goal is to identify the non-zero entries of $\btheta$ as well as estimate it. There are a wide variety of methods based on the penalized likelihood approach that solves the following optimization problem:

\begin{align}
  \min_{\btheta} \sum_{i=1}^{n} &  \left( y_i - \theta_0 - \sum_{j=1}^{p} \theta_j x_{i,j} \right)^2 + \text{pen}_{\lambda}(\btheta), \label{eq:penalize} \\
  \text{where } & \text{pen}_{\lambda}(\btheta) = \sum_{j=1}^{p} p_{\lambda}(\theta_j) \text{ is a separable penalty.} \nonumber
\end{align}

Lasso uses an $\ell_1$ penalty, $p_{\lambda}(\theta_j) = \lambda \abs{\theta_j}$, and simultaneously performs variable selection while maintaining estimation accuracy.  Another notable variant is the best subset selection procedure corresponding to the $\ell_0$ penalty $p_{\lambda}(\theta_j) = \lambda \1\{\theta_j \ne 0\}$. There has been a recent emphasis on non-concave separable penalties such as the minimax concave penalty or MCP \citep{zhang2010nearly} or SCAD \citep{fan2001variable}, that act as a tool for variable selection and estimation. 
We discuss the penalization methods from a Bayesian viewpoint in the next section.

\subsection{Variable Selection}

Variable or predictor selection is intimately related to high-dimensional sparse linear regression.  A sparse model provides interpretability, computational efficiency, and stability of inference. Lasso's success has inspired many estimation methods that rely on convexity and sparsity in a penalized estimation framework. The `bet on sparsity' principle \citep{hastie_elements_2009} dictates the use of methods favoring sparsity, as no method uniformly dominates when the true model is dense.  

\begin{remark}

The LAVA method by \citet{chernozhukov2017lava} strictly dominates both Lasso and ridge in a `sparse+dense' model. In fact, the LAVA estimator performs as well as Lasso in a sparse regime and as well as ridge \citep{tikhonov1963solution} in a dense regime. This questions the validity of the `bet on sparsity' principle. 
Although there is no exact analogue of LAVA in the Bayesian world, the one-group shrinkage priors share a common philosophy. The horseshoe-type priors are also designed to work when true $\btheta$ has a few large entries and very many small non-zero entries and produces a `non-sparse' estimator, but LAVA can recover both the dense and sparse components unlike horseshoe.

\end{remark}

A parallel surge of Bayesian methodologies has emerged for sparse regression problems with an underlying variable selection procedure. Hierarchical Bayesian modeling proceeds by selecting a model dimension $s$, selecting a random subset $S$ of dimension $\abs{S} = s$ and a prior $\pi_S$ on $\mathbb{R}^{s}$. The prior can be written as in \cite{castillo2015bayesian}:

\begin{equation}
  (S,\btheta) \mapsto \binom{p}{\abs{S}}^{-1} 
	\pi_p(\abs{S}) \pi_S(\btheta_S)\delta_{0}(\btheta_{S^c}). \label{eq:bayes-hier}
\end{equation}

\noindent Bayesian approaches for sparse linear regression include \citet{george2000variable, George0000, mitchell88, ishwaran2005spike} and more recently \cite{rovckova2016spike}, who introduce the spike-and-slab Lasso prior, where the hierarchical prior on the parameter and model spaces assumes the form:

\begin{equation}
  \pi(\btheta \mid \gamma) = \prod_{i=1}^{p} [\gamma_i \pi_1(\theta_i) +
  (1-\gamma_i) \pi_0(\theta_i)], \quad \gamma \sim p(\cdot), \label{eq:ssl}
\end{equation}

where $\bgamma$ indexes the $2^p$ possible models, and $\pi_0$, $\pi_1$ model the null and non-null $\theta_i$'s respectively using two Laplace priors with different scales. 
However, the spike-and-slab type priors lead to substantial computational challenges as exploring the full posterior using point mass mixture priors is prohibitive due to a combinatorial complexity of updating the discrete indicators and infeasibility of block updating of model parameters. 

The continuous shrinkage priors alleviate this by using efficient Gibbs sampling scheme based on block-updating the model parameters. We also note that while full posterior sampling remains a computational hurdle for the spike-and-slab prior, point estimates such as posterior mean and posterior quantiles can be obtained using a polynomial-time algorithm as shown by \citet{castillo2012needles}. \citet{rovckova2016spike} discuss the inefficiency of stochastic search algorithms for exploring the posterior even for moderate dimensions and developed a deterministic alternative to quickly find the maximum a-posteriori model. Here \begin{enumerate*}
  \item increasing the efficiency in computation in the spike-and-slab model remains an active area of research \citep[see, e.g., ][]{rovckova2016spike}
    and
  \item some complicating factors in the spike-and-slab model, such as a lack of suitable block updates, have fairly easy solutions for their continuous global-local shrinkage counterparts, facilitating posterior exploration. 
\end{enumerate*}

\citet{polson2010shrink, carvalho2010horseshoe, polson2012half} introduced the `global-local' shrinkage priors that adjust to sparsity via global shrinkage, and identify signals by local shrinkage parameters. The global-local shrinkage idea has resulted in many different priors in the recent past, with varying degrees of success in theoretical and numerical performance. We compare these different priors and introduce a recently proposed family of horseshoe-like priors in \S \ref{sec:one-gp}.   

The estimators resulting from the one-group shrinkage priors are very different from the shrinkage estimator due to \citet{james_estimation_1961}, who showed that maximum likelihood estimators for multivariate normal means are inadmissible beyond two dimensions. The James--Stein estimator is primarily concerned about the total squared error loss, without much regard for the individual estimates. In problems involving observations lying far away on the tails this leads to `over-shrinkage' \citep{carvalho2010horseshoe}. In reality, an ideal signal-recovery procedure should be robust to large signals. Connections between the global shrinkage of James--Stein and global-local shrinkage of the horseshoe are discussed in more details in \S \ref{sec:stat-prop}.


\section{Lasso and Horseshoe}\label{sec:3}


Regularization requires the researcher to specify a measure of fit, denoted by $l(\btheta)$ and a penalty function, denoted by $\text{pen}_{\lambda}(\btheta)$. From a Bayesian perspective,  $l(\theta)$ and $\text{pen}_{\lambda}(\btheta)$ correspond to the negative logarithms of the likelihood and a suitable prior distribution, respectively. While regularization leads to an optimization problem of the form 
\begin{equation}
  \min_{\btheta \in \mathbb{R}^p}
  \left\{ l(y \mid \btheta) + \text{pen}_{\lambda}(\btheta) \right\}
  \;, 
  \label{eq:reg}
\end{equation}
the probabilistic approach leads to a Bayesian hierarchical model
\begin{equation}
  p(y \mid \btheta) \propto \exp\{-l(y \mid \btheta)\} \; , \quad \pi_{\lambda}(\btheta)
  \propto \exp\{ -\text{pen}_{\lambda}(\btheta) \}. 
  \label{eq:pen}
\end{equation}

For appropriate $l(y \mid \btheta)$ and $\text{pen}_{\lambda}(\btheta)$, the solution to \eqref{eq:reg} corresponds to the posterior mode of \eqref{eq:pen}, $\hat{\btheta} = \argmax_{\btheta} p(\btheta \mid y)$, where $p(\btheta \mid y)$ denotes the posterior density. The properties of the penalty are then induced by those of the prior. For example, regression with a least squares log-likelihood subject to an $\ell_2$ penalty or ridge \citep{tikhonov1963solution, hoerl70} corresponds to a Gaussian prior under the same observation distribution, and an $\ell_1$ penalty (Lasso) corresponds to a double-exponential prior \citep{tibshirani96}.

One interpretation of Lasso and related $\ell_1$ penalties is that these are methods designed to perform selection, while ridge and related $\ell_2$ based methods perform shrinkage. Selection-based methods such as the Lasso are unstable in many situations, e.g., in presence of multi-collinearity in the design \citep[ch.3]{hastie_elements_2009}. 

Although `shrinkage' and `selection' are closely related, we tend to distinguish between them in the following sense. Shrinkage methods such as the horseshoe prior shrink towards $0$ by thresholding the shrinkage weights that behave like posterior inclusion probabilities $P(\theta_i \neq 0 \mid y_i)$ to achieve variable selection. It should be noted that the continuous nature of prior on $\theta_i$ ensures a lack of exact zeros in the posterior, which is often preferred over dichotomous models by some practitioners \citep{stephens2009bayesian} as more realistic.  This is unlike the Lasso that performs explicit selection by making some of estimates $0$ and producing a true sparse solution. Ultimately, both selection and shrinkage have their advantages and disadvantages. 


\subsection{Lasso Penalty and Prior}\label{sec:lasso}

%

As discussed before, the classical Lasso-based point estimate is the same as the posterior mode under component-wise Laplace prior, and the mode inherits the optimal properties of Lasso. For example, the oracle inequality in \citet[Eq. (2.8), Th. (6.1)]{buhlmann2011statistics} states that with a proper choice of $\lambda$ of order $\sigma \sqrt{\log(p)/n}$, the mean squared prediction error of Lasso is of the same order as if one knew active set $S_0 = \{j : \theta_j^0 \neq 0 \}$), up to $O(\log(p))$ and a compatibility constant $\phi_0^2$. The compatibility (or restricted eigenvalue) constant reflects the compatibility between the design matrix and the $\ell_1$ norm of $\btheta$, and is defined as follows \citep[Eq (6.4)]{buhlmann2011statistics}:

\begin{definition}\label{def:compatibility}(Compatibility Condition). 
For $S \subset \{1,2,\ldots,p\}$ and $\btheta \in \mathbb{R}^{p}$, let $\btheta_{j,S} \doteq \theta_j 1\{j \in S\} \in \mathbb{R}^{p}$ (with similar notation for $\btheta_{j \in S} \in \mathbb{R}^{\abs{S}}$), and let $\btheta_{-S} = \btheta_{S^c}$. Then the compatibility condition is satisfied for the design $\X$ for the true support set $S = \text{supp}(\btheta)$, if letting $s_0 = \abs{S}$ one has,
\[
\frac{1}{n} \vectornorm{\X\btheta}_2^2 \ge \frac{\phi_0^2}{s_0} \vectornorm{\btheta_{S}}_1^2, \; \text{ for all } \; \btheta \in \mathbb{R}^p \; \text{such that} \; \vectornorm{\btheta_S}_1 \le 3 \vectornorm{\btheta_{-S}}_1.
\]
The constant $\phi_0^2$ is called the compatibility (or restricted eigenvalue) constant.
\end{definition}

Lasso also exhibits other desirable properties such as computational tractability, consistency of point estimates of $\btheta$ for suitable $\lambda$, and optimality results on variable selection.

\subsection{Bayesian Lasso and Elastic Net}

As discussed before, the posterior mean under the double-exponential prior, which is the Bayes estimate under squared error loss, does not satisfy the optimality properties of the posterior mode under the double-exponential prior (i.e., the Lasso). Along these lines, \citet{castillo2015bayesian} argue that the Lasso is essentially non-Bayesian, in that the ``\textsl{full posterior distribution is useless for uncertainty quantification, the central idea of Bayesian inference}." \citet{castillo2015bayesian} provide theoretical result that the full Lasso posterior does not contract at the same speed as the posterior mode. 

Thus, there are a number of caveats related to the use of a double-exponential prior for the general purposes of shrinkage.  An important example is found in how it handles shrinkage for small observations and robustness to the large ones. This behavior is described by various authors, including \cite{polson2010shrink,datta2013asymptotic}, and motivates the key properties of global-local priors. Figure~\ref{fig:priorkappa} provides profile plots as a diagnostic of shrinkage behavior for different priors.

For correlated predictors, \cite{zou2005regularization} proposed a family of convex penalties called `elastic net', which is a hybrid between Lasso and ridge. The penalty term is $\sum_{j=1}^{p} \lambda p_{\alpha}(\theta_j)$, where 

\[
p_{\alpha}(\theta_j) = \half (1-\alpha)\theta_j^2 + \alpha \abs{\theta_j}, \quad j = 1, \ldots, p. 
\]

Both Lasso and elastic net facilitate efficient Bayesian computation via a global-local scale mixture representation \citep{bhadra2016global}. The Lasso penalty arises as a Laplace global-local mixture \citep[][]{andrews1974scale}, while the elastic-net regression can be recast as a global-local mixture with a mixing density belonging to the orthant-normal family of distributions
\citep{hans2011elastic}.  The orthant-normal prior on $\theta_i$, given hyper-parameters $\lambda_1$ and $\lambda_2$, has a density function with the following form:

\begin{equation}
  p(\theta_i \mid \lambda_1, \lambda_2)  = 
  \begin{cases} 
   \phi(\theta_i \mid \frac{\lambda_1}{2\lambda_2}, \frac{\sigma^2}{\lambda_2}) / 2\Phi\left(-\frac{\lambda_1}{2\sigma \lambda_2^{1/2} }\right), & \quad \theta_i < 0, \\
   \phi(\theta_i \mid \frac{-\lambda_1}{2\lambda_2}, \frac{\sigma^2}{\lambda_2}) / 2\Phi\left(-\frac{\lambda_1}{2\sigma \lambda_2^{1/2} }\right), & \quad \theta_i \geq 0. \end{cases} 
  \label{eq:hans}
\end{equation}

\subsection{Horseshoe Penalty and Prior}\label{sec:one-gp}

The horseshoe prior is a continuous shrinkage rule for sparse signal recovery. Here we discuss the motivation behind the horseshoe prior for the Gaussian sequence model as it was developed in \citet{carvalho2010horseshoe}, but note that it is applicable to sparse signal recovery in regression models and beyond, as we discuss in \S \ref{sec:sparse-linreg}.  Consider the normal means model: $y_i \mid \theta_i \sim \Nor(\theta_i,1), \theta_i \mid \lambda_i, \tau \sim \Nor(0, \lambda_i^2 \tau^2)$, $i = 1,2, \ldots, n$. The horseshoe prior for $\theta_i$, given a global shrinkage parameter $\tau$, is given by the hierarchical model 

\begin{align}
  (y_i \mid \theta_i) & \sim \Nor(\theta_i , \sigma^2),\;  (\theta_i \mid \lambda_i, \tau) \sim  \Nor(0, \lambda_i^2 \tau^2), \nonumber \\
 \lambda_i ^2 &  \sim \operatorname{C}^{+}(0,1), \quad i = 1, \ldots, n.  \label{eq:hs}
\end{align}

As discussed before, the spike-and-slab prior or the two-groups model (\textit{vide} Appendix \ref{sec:2gp}) with two dedicated components for separating noise and signal is a natural Bayesian solution but it leads to substantial computational burden. The horseshoe prior takes a different approach: instead of placing a prior on the model space to yield a sparse estimator, it models the posterior inclusion probabilities $P(\theta_i \ne 0 \mid y_i)$ directly. 
To see this, note that the posterior mean under the horseshoe prior can be written as a linear function of the observation:

\beq
\E(\theta_i \mid y_i) = \{1- \E(\kappa_i \mid y_i) \} y_i \text{ where } \kappa_i = 1/(1+\lambda_i^2 \tau^2).
\eeq

The name `horseshoe' arises from the shape of the beta prior density of the shrinkage weights $\kappa_i$. A comparison with the posterior mean obtained under the two-groups model reveals that the shrinkage weights perform the same function as the posterior inclusion probability $P(\theta_i \ne 0 \mid y_i)$ for recovering a sparse signal. Since the shrinkage coefficients are not formal Bayesian posterior quantities, we refer to them as `pseudo posterior inclusion probabilities.' 

\citet{carvalho2010horseshoe} provided strong numerical evidence that the shrinkage weights from a one-group prior accurately approximates the inclusion probabilities under a two-groups model, and used this property to construct a multiple testing rule. The thresholding rule rejects the $i^{th}$ null hypothesis $H_{0i}: \theta_i = 0$ if the shrinkage weight $1-\hat{\kappa}_i$ exceed $0.5$. \citet{datta2013asymptotic} validated this theoretically by proving that the horseshoe multiple testing rule attains the Bayes oracle up to a multiplicative constant under a $0$-$1$ additive loss. 

The marginal likelihood after reparametrizing $\kappa_i = (1+\lambda_i^2\tau^2)^{-1}$ is:  $p(y_i \mid \kappa_i, \tau) = \kappa_i^{1/2} \exp \left(-\kappa_i y_i^2/2 \right)$. The posterior density of $\kappa_i$ identifies signals and noises by letting $\kappa_i \to 0$ and $\kappa_i \to 1$ respectively. Since the marginal likelihood is zero when $\kappa_i = 0$, it does not help identify the signals. Intuitively, any prior that drives the probability to either extremities should be a good candidate for sparse signal reconstruction. The horseshoe prior, with an induced prior density on $\kappa_i$ proportional to $\kappa_i^{-1/2} (1-\kappa_i)^{-1/2}$  does exactly that: it cancels the $\kappa_i^{1/2}$ term in the marginal likelihood and replaces it with $(1-\kappa_i)^{-1/2}$ to enable $\kappa_i \to 1$ in the posterior. The horseshoe+ prior \citep{bhadra2015horseshoe+} takes this philosophy one step further, by creating a $U$-shaped Jacobian for transformation from $\lambda_i$ to $\kappa_i$-scale. The double-exponential on the other hand, yields a prior that decays at both ends with a mode near $\kappa_i = 1/4$, thus leading to a posterior that is neither good at adjusting to sparsity, nor at recovering large signals.
 
\begin{figure}[!ht]
\begin{subfigure}{0.48\linewidth}
\centering
\includegraphics[height = 2in, width=\textwidth]{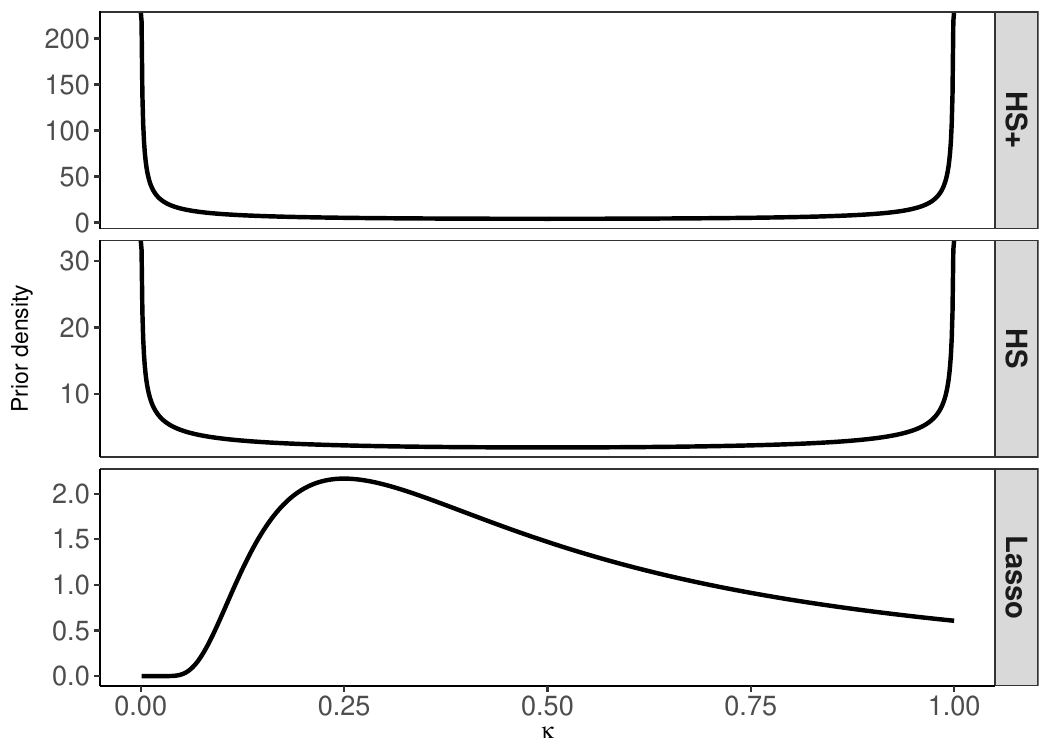}
\caption{\footnotesize{Shrinkage factor $\kappa_i$ against $p(\kappa_i)$}}
\label{fig:priorkappa}
\end{subfigure}
\begin{subfigure}{0.48\linewidth}%
\centering
\includegraphics[height = 2in, width=\columnwidth]{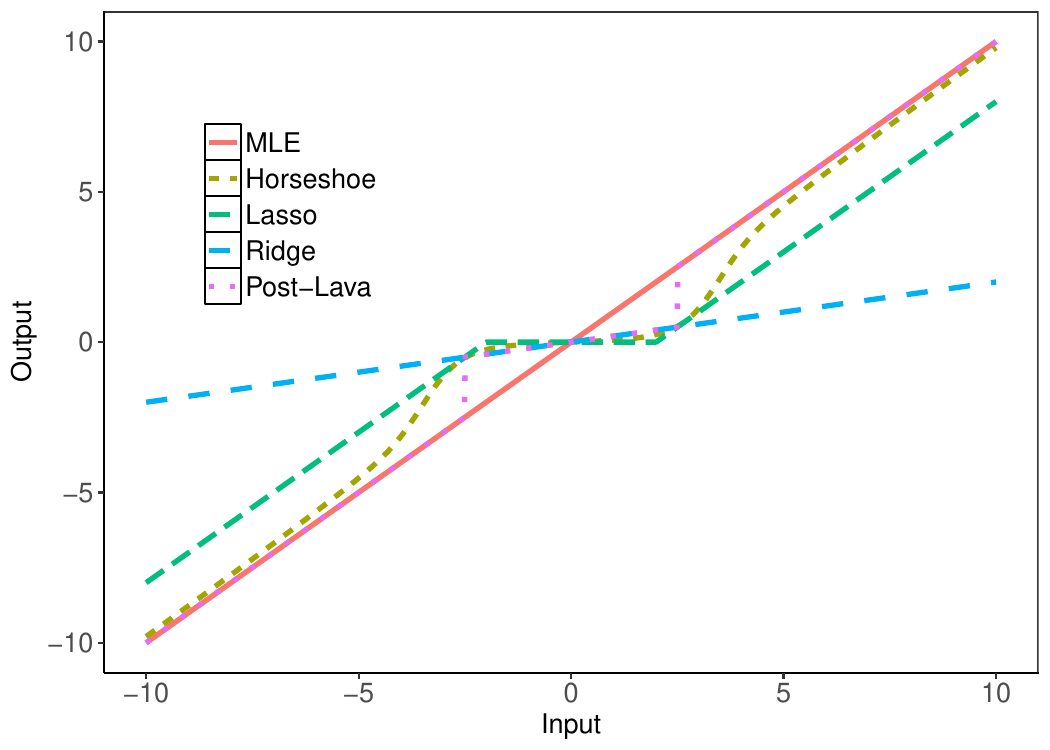}%
\caption{Score function}%
\label{fig:score}%
\end{subfigure}
\label{fig:score-kappa}
\caption{\footnotesize{(a) Prior density of shrinkage weight $\kappa_i$ for the horseshoe, horseshoe+, and Laplace
prior, where $\{1- E(\kappa_i \mid y_i)\}$ can be interpreted as the pseudo posterior inclusion probability that mimics $P(\theta_i \ne 0 \mid y_i)$, and (b) shrinkage function for LAVA, Lasso, ridge and the horseshoe estimator. For
LAVA shrinkage function, we have chosen $\lambda_1 = \lambda_l = 4$ and $\lambda_2 = \lambda_r = 4$, and for the horseshoe prior the value of global shrinkage parameter $\tau$ is fixed at $0.1$.}}
\end{figure}

Figure \ref{fig:priorkappa} plots the prior density $p(\kappa_i)$ for the horseshoe, horseshoe+, and the Laplace priors. Figure \ref{fig:score} shows the resulting shrinkage function by plotting the input observations against the output estimates for horseshoe, horseshoe+, and Laplace priors, along with the maximum likelihood estimator ($\hat{\btheta} = \y$). Both Lasso and horseshoe shrink the small observations, but while horseshoe and horseshoe+ leave the large inputs unshrunk, Lasso shrinks them by a non-vanishing amount, resulting in a non-zero bias. We also plot the shrinkage function for the post-lava estimator \citep{chernozhukov2017lava} (\textit{vide} Appendix C) which works well on dense+sparse signals, and has the robustness property lacking in Bayesian Lasso or the Laplace prior. 

There are a number of closed-form results for the posterior distribution under a horseshoe prior. Although the prior density under the horseshoe prior does not admit a closed form, we can write the horseshoe posterior mean using the Tweedies' formula $\E(\theta \mid y) = y + \dd{\ln m(y)}{y}\sigma^2$, which is also the Bayes adjustment that provides an optimal bias-variance trade-off. For the horseshoe prior, Tweedies' formula yields:

\begin{equation}
  \E(\theta_i \mid y_i, \tau) = y_i \left( 1 - \frac{2\Phi_1(\half, 1,
  \frac{5}{2},\frac{y_i^2}{2\sigma^2}, 1-\frac{1}{\tau^2})}{
  3\Phi_1(\half, 1, \frac{3}{2},\frac{y_i^2}{2\sigma^2}, 1-\frac{1}{\tau^2})} \right)
  \;,
\end{equation}
where $\Phi_1$ is the bivariate confluent hypergeometric function \citep{gordy1998computationally}. A similar formula is available for the posterior variance. This enables one to rapidly calculate the posterior mean estimator under the horseshoe prior via a `plug-in' approach with estimated values of the hyper-parameter $\tau$. 
We discuss the different approaches for handling $\tau$ in \S \ref{sec:5} and statistical properties of horseshoe posterior mean estimator and the induced decision rule in more details in \S~\ref{sec:stat-prop}. 

The horseshoe prior is a member of a wider class of global-local scale mixtures of normals that admit following hierarchical form \citep{polson2010shrink}: 

\begin{gather*}
(\y \mid \btheta) \sim \Nor(\btheta, \sigma^2 \I) ;\; \theta_i \sim \Nor(0, \lambda_i^2 \tau^2), \\
\lambda_i^2 \sim \pi(\lambda_i^2) ; \; (\tau,\sigma^2) \sim  \pi(\tau^2,\sigma^2), i = 1, \ldots, n. 
\end{gather*}

These priors are collectively called global-local shrinkage priors in \cite{polson2010shrink}, since they recover signals by a local shrinkage parameter and adapt to sparsity by a global shrinkage parameter. Table \ref{tab:one-gps} provides a list of the popular and recent global-local shrinkage priors. 
A natural question is \textit{how do we compare these priors?} It is known due to several authors \citep[e.g., ][]{polson2010shrink,bhadra2015default} that the key features of a global-local shrinkage prior is a peak at origin and heavy tails. An early example of such a prior was proposed by \citet{cutillo2008larger} in the context of wavelet thresholding where a heavier tail was attained by modeling $\theta \sim \Nor(0, \tau^2)$ and $\tau^2 \sim (\tau^2)^{-k}$ where $k > 1/2$. 
We list a few popular global-local shrinkage priors along with their behavior near origin and the tails on Table \ref{tab:one-gps}, and a discussion of the recent extensions of global-local priors beyond the Gaussian model is deferred to \S~\ref{sec:app-ext}.

\begin{table}%
\centering
\caption{Origin and tail behaviors of different priors}
\label{tab:priors}
\begin{tabular}{| c | c |c |}
\hline
Prior & Origin Behavior & Tails \\
\hline 
Horseshoe & $-\log(\abs{\theta})$ & $\abs{\theta}^{-2}$ \\
Horseshoe+ & $-\log(\abs{\theta})$ & $\abs{\theta}^{-1}$ \\
Horseshoe-like & $-\abs{\theta}^{1-\epsilon}\log(\abs{\theta})$ & $\abs{\theta}^{1-\epsilon}$ $\epsilon \ge 0$\\
GDP & Bounded at origin & $\abs{\theta}^{-(\alpha + 1)}, \alpha \ge 0$ \\
$DL_{a}$ ($DL_{\frac{1}{n}}$) & $\abs{\theta}^{a-1}$ ($\abs{\theta}^{\frac{1}{n}-1}$) & $\exp(-b\abs{\theta})$ \\
\hline
\end{tabular}
\end{table}

\begin{figure}[ht!]
  \begin{subfigure}{0.45\linewidth}
	\includegraphics[height=2.5in,width=\textwidth]{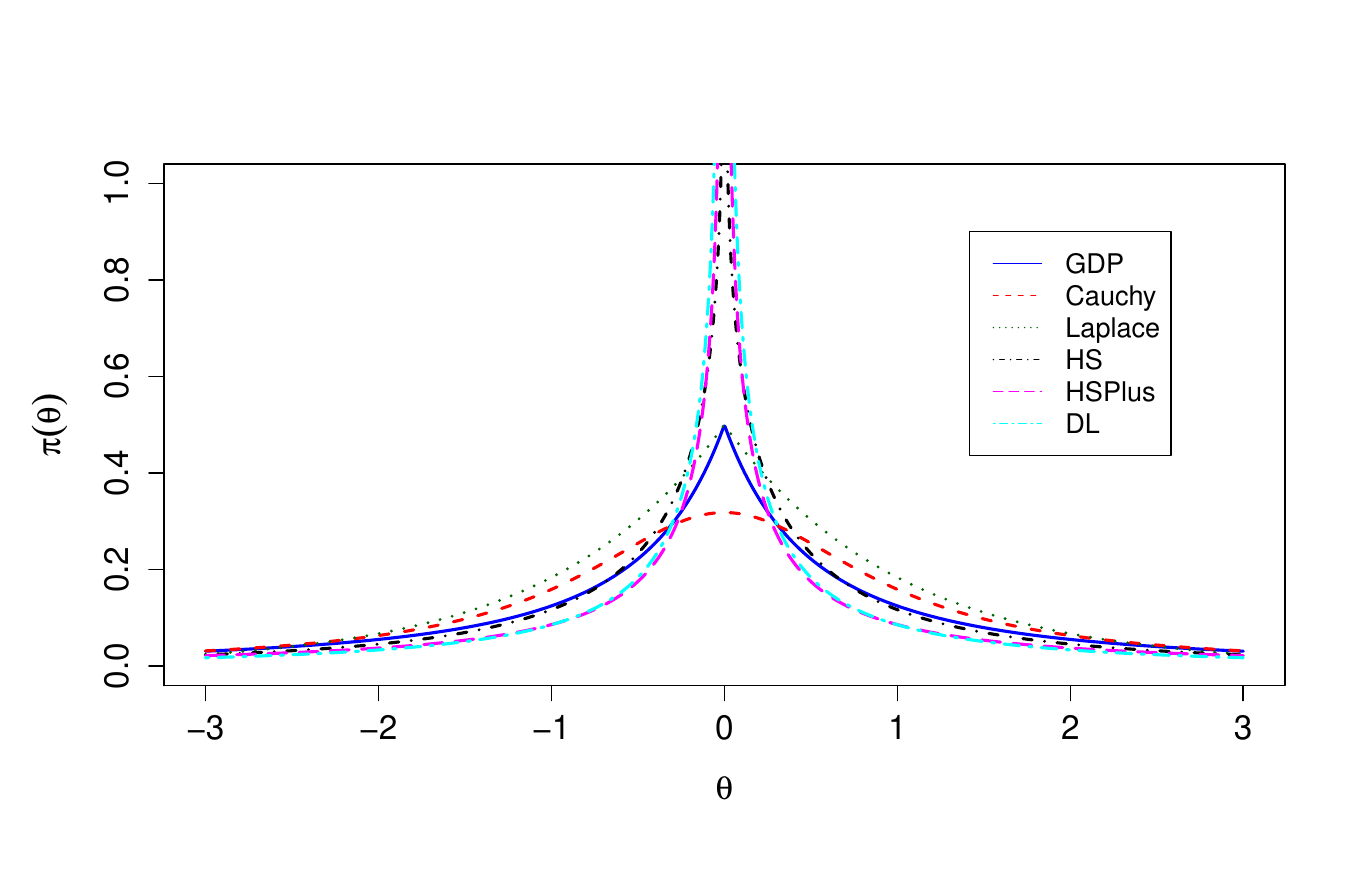}%
	\label{fig:zero}
	\end{subfigure}
	\hspace{0.1in}
  \begin{subfigure}{0.45\linewidth}
	\includegraphics[height=2.5in,width=\textwidth]{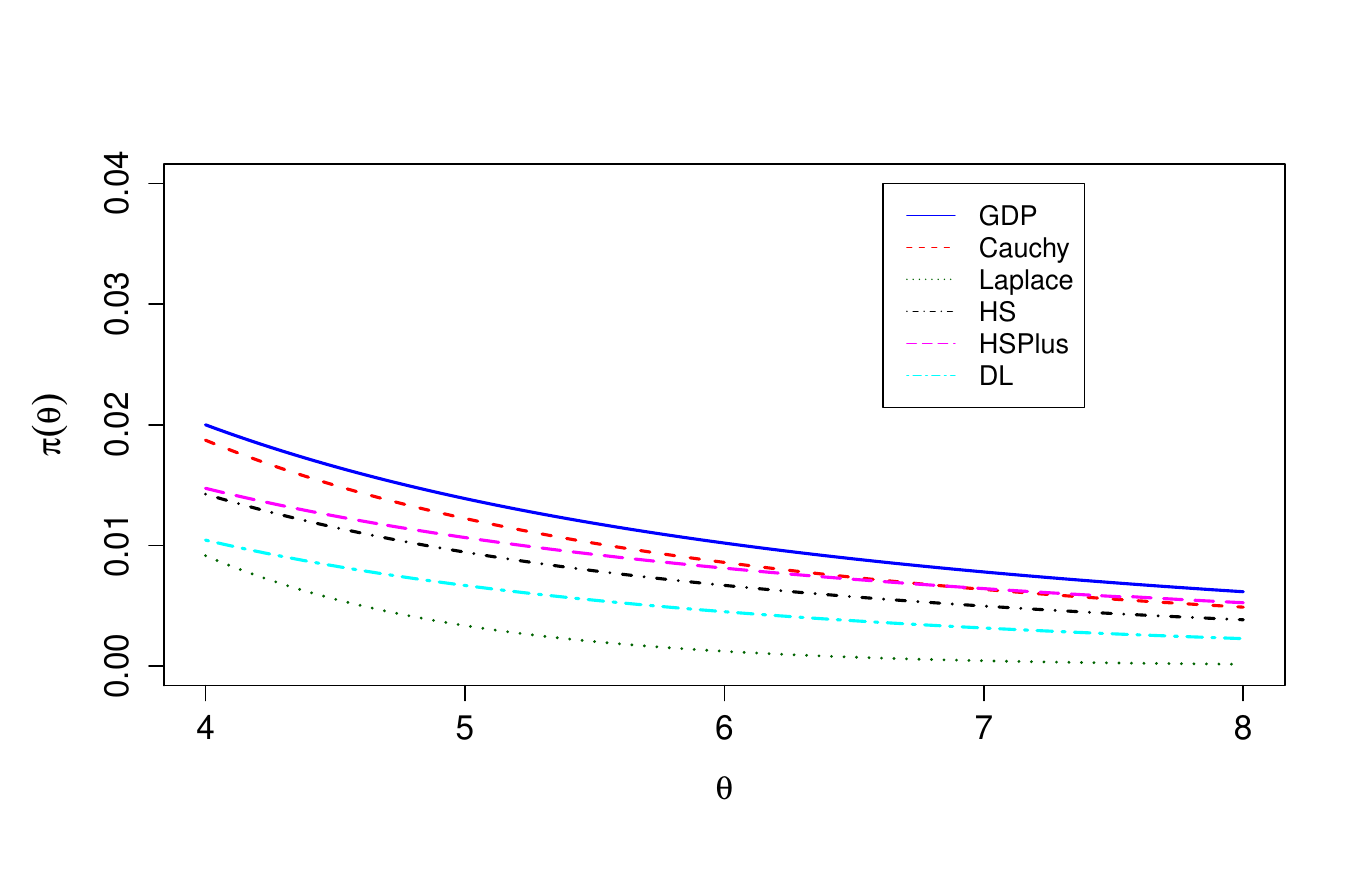}
  \label{fig:tails}
		\end{subfigure}
 \caption{Marginal prior densities near the origin (left) and in the tail regions (right). The legends denote the horseshoe+ (HSPlus), horseshoe (HS), Dirichlet-Laplace (DL), generalized double Pareto (GDP), Cauchy and Laplace priors.}
\end{figure}

From a regularization view-point, one way to judge a prior is by the penalty it imposes on a likelihood \eqref{eq:reg}, although in a strict Bayesian spirit, a prior should be evaluated based on the whole posterior, as shown by several authors including \citet{castillo2015bayesian} and \citet{van2017adaptive}. %
Although the horseshoe prior leads to optimal performance as a shrinkage prior, the induced penalty $\log \pi(\theta)$ does not admit a closed form as the marginal prior $\pi(\theta)$ is not analytically tractable. This poses a hindrance in learning via Expectation-Maximization or other similar algorithms.  The generalized double Pareto prior of
\citet{armagan2011generalized} admits a closed form solution, but it does not have an infinite spike near zero needed for sparse recovery. Motivated by this fact, \citet{bhadra2017horseshoe} recently proposed the `horseshoe-like' prior by normalizing the tight bounds for the horseshoe prior. Thus, the horseshoe-like prior attains a unique status within its class: it has a closed form marginal prior for $\theta$, yet with a spike at origin and heavy tails and more importantly, admits a global-local scale mixture representation. The scale mixture representation supports both a traditional MCMC sampling for uncertainty
quantification in full Bayes inference and EM/MM or proximal learning when computational efficiency is the primary concern. 

Since the aim of designing a sparsity prior is achieving higher spike near zero while maintaining regularly varying tails, a useful strategy is to split the range of the prior into disjoint intervals: $[0,1)$ and $[1, \infty)$, and aim for higher spike in one and heavier tail in the other. This leads to a class of `horseshoe-like' priors with more flexibility in shape than any single shrinkage prior. We provide the general form of horseshoe-like priors and a key representation theorem.  The proof that horseshoe-like prior is a scale mixture with Slash normal mixing density involves Frullani's probabilistic identity \citep[\textit{vide}][pages 406-407]{jeffreys1972methods}, and to save substantial additional space we refer the readers to the proof in \S 5, Lemma 5.1 and Proposition 5.1 of \citet{bhadra2017horseshoe}.

\begin{description}
  \item[Horseshoe-like priors:] 
    \citet{bhadra2017horseshoe} have the following marginal prior density for
    $\theta_i$: 
    \begin{equation}
      \tilde p_{\tilde{HS}} (\theta_i \mid \tau^2) = \frac{1}{2 \pi{\tau}}\log
      \left ( 1 + \frac{\tau^2}{\theta_i^2} \right ), \quad  \; \theta_i  \in
      \mathbb{R},\; \tau > 0. \label{eq:hslike}
    \end{equation}
    The general family of horseshoe-like priors can be constructed as a density
    split into disjoint intervals as follows:
    \begin{align}
      p_{hs}(\theta_i \mid \tau^2)&  \propto \begin{cases}
        \frac{1}{\theta_i^{1-\epsilon}}\log \left( 1 + \frac{\tau^2}{\theta_i^2} \right) & \text{ if } {\abs{\theta_i} < 1}, \\ 
\theta_i^{1-\epsilon} \log \left( 1 + \frac{\tau^2}{\theta_i^2} \right) & \text{ if }{\abs{\theta_i} \ge 1}, \\ \end{cases} \; \epsilon \ge 0,\tau > 0.
        \label{eq:split} 
    \end{align}
  \item[Normal scale mixture:] 
    The horseshoe-like prior \eqref{eq:hslike} is a Gaussian scale mixture with  a Slash Normal mixing density, which is in turn another Gaussian scale mixture of $\operatorname{Pareto}(1/2)$ density, yielding the following representation theorem: 
    \begin{theorem}\citep{bhadra2017horseshoe}.\label{th:hslike}
      The horseshoe-like prior in \eqref{eq:hslike} has the following global-local scale mixture representation:
      \begin{equation}
        \begin{gathered}
    (\theta_i \mid t_i, \tau) \sim \Nor\left(0, \frac{\tau^2}{t_i^2} \right),\quad (t_i \mid s_i) \sim \Nor\left(0, s_i \right),\\
          s_i \sim \operatorname{Pareto}\left( \frac{1}{2} \right), \quad t_i \in \mathbb{R}, \; \tau \ge 0.
  \end{gathered}\label{eq:pareto}
      \end{equation}
    \end{theorem}
\end{description}

\begin{table}[!ht]
\centering
\caption{Priors for $\lambda_i$ and $\kappa_i$ for a few popular shrinkage rules}
\begin{tabular}{ccc}
\hline
Prior for $\theta_i$ & Prior for $\lambda_i$ & Prior for $\kappa_i$ \\ 
\hline \\
Horseshoe & $2/ \left\{ \pi \tau (1 + (\lambda_i/\tau)^2 )\right\}$  & $\frac{\tau}{\sqrt{\kappa_i (1-\kappa_i )}} \frac{1}{(1+\kappa_i (\tau^2 -1 ) )}$ \\[10pt]
Horseshoe+ & $\frac{4\log \lambda_i/\tau}{\left\{{\pi^2 \tau}(\lambda_i/\tau)^2 -1)\right\}}$ &  $\frac{\tau}{\sqrt{\kappa_i (1-\kappa_i )}}\frac{\log \left \{ ( 1 - \kappa_i ) / \kappa_i \tau^2 \right \}}{ (1-\kappa_i (\tau^2 +1 ))}$ \\[10pt]
Double Exponential & $\lambda_i \exp (-\lambda_i^2/2)$ & $\kappa_i^{-2} \exp{-\frac{1}{2\kappa_i}}$ \\
\hline 
\end{tabular}
\end{table}

\section{Statistical Risk Properties}\label{sec:stat-prop}
\subsection{Inadmissibility of MLE} 

We briefly discuss the Stein shrinkage phenomenon as it provides an useful insight into the development of global-local shrinkage estimators in high-dimensional problems. The James--Stein (JS) estimator is $\estJs(\y) = \{1 - (n-2)/\vectornorm{\y}^2 \} \y$ which is equivalent to the empirical Bayes estimate $\hat{\theta}_{\mathrm{Bayes}} = \hat{\tau}^2/(\hat{\tau}^2+1)\y$, under i.i.d. $\Nor(0,\tau^2)$ priors on $\theta_i$ and $\hat{\tau}$ being the empirical Bayes estimate of $\tau$ from the data \citep{efron2010large}. Thus, the James--Stein estimator corresponds to the Bayes risk of $n{\tau^2}/({\tau^2+1}) + 2/(1+\tau^2)$.
We argue below that a global shrinkage rule such as the James--Stein estimator or $\ell_2$ regularization does not work in the sparse regime as it lacks local parameters for handling sparsity. 

The story of shrinkage estimation goes back to the proof in \citet{stein_inadmissibility_1956} that the maximum likelihood estimators for normal data are inadmissible beyond $\mathbb{R}^2$. \citet{james_estimation_1961} proved that this estimator dominates the MLE in terms of the expected total squared error for every choice of $\btheta$, i.e. it outperforms the MLE no matter what the true $\btheta$ is. To motivate the need for developing a local shrinkage rule, consider the classic James--Stein (JS) `global' shrinkage rule, $\estJs(\y)$. The JS estimator uniformly dominates the traditional sample mean estimator, $\hat{\btheta}$. For all values of the true parameter $\btheta$ and for $n>2$, we have the classical mean squared error (MSE) risk bound:

\[ 
  R(\estJs, \btheta) \defeq \E_{y \mid \btheta} {\Vert \estJs(\y) - \btheta
  \Vert}^2 < n = \E_{\y \mid \btheta} {\Vert \y - \btheta \Vert}^2, \quad
  \forall \btheta \in \mathbb{R}^n, \; n \ge 3.  
\]

For sparse signal problem the standard James--Stein shrinkage rule, $\estJs$, performs poorly. This is best seen in the sparse setting for a $r$-spike parameter value $ \theta_r$ with $r$ coordinates at $
\sqrt{n/r} $ which has $ \Vert \theta \Vert^2 =n $. \citet{johnstone2004needles} show that $E \Vert \estJs - \btheta \Vert \leq n $ with risk $2$ at the origin. This leads to a bound (for $\sigma^2 = 1$)

\[
  \frac{n \Vert \btheta \Vert^2}{ n + \Vert \btheta \Vert^2} \leq R \left (
  \estJs, \theta_r \right ) \leq 2 + \frac{n \Vert \btheta \Vert^2}{ n
  + \Vert \btheta \Vert^2},
\]

The lower bound is the risk of an `ideal' linear estimator $\hat{\btheta}_c(\y) = c\y$. For an `ideal' estimator, $\vectornorm{\btheta}$ is known and $c$ is chosen to minimize the MSE, which gives 

\beq 
\tilde{c}(\btheta) = \vectornorm{\btheta}^2 / (n + \vectornorm{\btheta}^2). \label{eq:ideal}
\eeq

Theorem 5 of \cite{donoho1995adapting} states the following result, an \textit{oracle inequality} for the James--Stein estimator: 

\begin{lemma}
Consider the `ideal' estimator $\tilde{\btheta}_{IS}(\y) = \tilde{c}(\btheta)(\y)$ in \eqref{eq:ideal}. For all $p \ge 2$ and for all $\btheta \in \mathbb{R}^p$, 

\[
R(\estJs(\y), \btheta_r) \le 2 + \inf_c R(\hat{\btheta}_c(\y), \btheta_r) = 2 + R(\tilde{\btheta}_{IS}(\y), \btheta_r).
\]

\end{lemma}

Here, $\estJs(\y)$ for the $r$-spike parameter value has risk at least $R\left( \hat{\theta}^{JS} , \theta_r \right) \geq (n/2)$. This is nowhere near optimal. As \citet{donoho1994ideal} showed, simpler rules such as the hard-thresholding and soft-thresholding estimates given by $\hat{\btheta}^{H}(\y,\lambda) = \y I\{\abs{\y} \ge \lambda \}$ and
$\hat{\btheta}^{S}(\y,\lambda) = sgn(\y) (\abs{\y} - \lambda)_{+}$ satisfy an oracle inequality. In particular, when the thresholding sequence is close to
$\sqrt{2\log n}$ (universal threshold), these estimators attain the `oracle risk' up to a factor of $2\log(n)$. Intuitively, this is not
surprising as the high-dimensional normal prior places most of its mass on circular regions -- and does not support sparse, spiky vectors. 
The James--Stein estimator was not built for sparse estimation and it is ambivalent to sparsity assumptions, but the shrinkage phenomenon in lower dimensional regime paves the way for building shrinkage rules for sparse regime, where one needs an additional `local' shrinkage term to recover the signals.

\subsection{Near minimax risk}

The asymptotically minimax risk rate in $\ell_2$ for nearly black objects is given by \citet{donoho1992maximum} to be $p_n \log \left ( n / p_n \right )$. Here $a_n \asymp b_n$ means $\lim_{n\to\infty} a_n/b_n=1$. Specifically, for any estimator $\delta(\y)$, we have a lower bound ($\sigma^2  = 1$): 

\begin{equation}
  \sup_{\theta_0 \in \ell_0[p_n]} \E_{\theta_0} \norm{\delta(Y) - \theta_0}^2
  \ge 2 p_n \log(n/p_n)(1+o(1)). 
\end{equation}

The minimax rate, which is a frequentist criteria for evaluating the convergence of point estimators to the underlying true parameter, is a validation criteria for posterior contraction as well. This result, due to \citet{ghosal2000}, showed that the minimax rate is the fastest that the posterior distribution can contract. 

A key advantage of the horseshoe estimators is that they enjoy near-minimax rates in both an empirical Bayes and full Bayes approach, provided that the hyper-parameters or the priors are suitably chosen--as proved in a series of papers \citep{van2014horseshoe,van2015conditions,van2016many,van2017adaptive}. Specifically, for $\sigma^2 = 1$, the horseshoe estimator achieves

\begin{equation}
  \sup_{ \btheta \in \ell_0[p_n] } \; \mathbb{E}_{ \y \mid \btheta } \norm{ \estHs (\y) -
  \btheta }^2 \asymp p_n \log \left ( n / p_n \right ),
  \label{eq:minimax}
\end{equation}

\citet{van2014horseshoe} showed that the near-minimax rate can be achieved by setting the global shrinkage parameter $\tau = (p_n/n) \log(n/p_n)$. In practice, $\tau$ is unknown and must either be estimated from the data or handled via a fully Bayesian approach by putting a suitable prior on $\tau$. \cite{van2017adaptive} show that the theoretical optimality properties for the popular horseshoe prior holds true if the global shrinkage parameter $\tau$ is learned via the maximum marginal likelihood estimator (MMLE) or a full Bayes approach. 
Independently, \citet{van2015conditions} and \citet{ghosh2016asymptotic} showed that these optimality properties are not unique features of the horseshoe prior and they hold for a general class of global-local shrinkage priors. While the results of \cite{van2015conditions} apply to a wider class of priors, including the horseshoe+ prior \citep{bhadra2015horseshoe+} and spike-and-slab Lasso \citep{rovckova2016spike}, it is worth pointing out the difference between \citet{van2015conditions} and \citet{ghosh2016asymptotic}. \citet{van2015conditions} prove `near-minimaxity' under `uniform regular variation' conditions on the prior on local shrinkage parameters for a general class of global-local priors that allow exponential tails. On the other hand, \citet{ghosh2016asymptotic} attain `exact' minimaxity for `horseshoe-type' priors under suitable conditions on the global parameter $\tau$, but they allow only polynomial tails, leading to a narrower class.

\subsection{Variable Selection: Frequentist and Bayes Optimality}

Here we compare the relative performance of horseshoe and Lasso for multiple testing under the two-groups model and a $0$-$1$ additive loss framework. One of the main reasons behind the widespread popularity of Lasso is the in-built mechanism for performing simultaneous shrinkage and selection. 
The horseshoe estimator, on the other hand, is a shrinkage rule that induces a selection rule through thresholding the pseudo posterior inclusion probabilities. \citet{datta2013asymptotic} proved that for large scale testing problems the horseshoe prior attains the oracle property while double-exponential tails prove to be insufficiently heavy, leading to a higher misclassification rate compared to the horseshoe prior. The main reasons behind the horseshoe prior's optimality are the posterior density of shrinkage weights that concentrates near $0$ and $1$ and the adaptability of the global shrinkage parameter $\tau$. 

The posterior distribution under the horseshoe prior leads to a natural model selection strategy under the two-groups model. \citet{carvalho2010horseshoe} argued that the shrinkage coefficient $1-\hat{\kappa}_i$ can be viewed as a pseudo-inclusion probability $P(\theta_i \ne 0 \mid y_i)$ and induces a multiple testing rule: 

\begin{equation}
  \text{Reject the $i^{th}$ null hypothesis } H_{0i} : \theta_i = 0 \text{ if }
  1-\hat{\kappa}_i > \half \;. 
  \label{eq:hsrule}
\end{equation}
Under the two-groups model \eqref{twogroups}, and a $0$-$1$ loss, the Bayes risk is 
\[
R = \sum_{i=1}^{n} \{ (1- \pi) t_{1i} + \pi t_{2i} \}, 
\]
where $t_{1i}$ and $t_{2i}$ denote the probabilities of type 1 and type 2 error corresponding to the $i^{th}$ hypothesis respectively. 

If we know the true proportion of sparsity and the parameters of the non-null distribution, we can derive a decision rule that is impossible to outperform in theory, which is called the Bayes oracle for multiple testing \citep{bogdan2011asymptotic}. The oracle risk serves as the lower bound for any multiple testing rule under the two-groups model and thus provides an asymptotic optimality criteria when the number of tests go to infinity. The framework of \citet{bogdan2011asymptotic} is: 
\begin{equation}
\pi_n \to 0, \; u_n = \psi_n^2 \to \infty, \; \text{and} \; \log(v_n)/u_n \to C \in (0,\infty) \label{eq:asymp}
\end{equation}
where $v_n = \psi_n^2 (\frac{1-\pi_n}{\pi_n})^2$. Dropping the subscript $n$ from parameters for notational simplicity, the Bayes risk for the Bayes oracle under the above framework \eqref{eq:asymp} is:
\[
R_{\text{oracle}} = n \pi (2 \Phi(\sqrt{C}) - 1)(1+o(1)).
\]
A multiple testing rule is said to possess asymptotic Bayes optimality under sparsity (ABOS) if it attains the oracle risk as $n \to \infty$. \citet{bogdan2011asymptotic} provided conditions for a few popular testing rules, e.g. Benjamini--Hochberg FDR controlling rule to be ABOS. \citet{datta2013asymptotic} first showed that the horseshoe decision rule \eqref{eq:hsrule} is also ABOS up to a multiplicative constant if $\tau$ is chosen suitably to reflect the sparsity, namely $\tau = O(\pi)$. The proof in \citet{datta2013asymptotic} hinges on the concentration of the posterior distribution near $0$ and $1$, depending on the trade-off between signal strength and sparsity.  In numerical experiments, \citet{datta2013asymptotic} also confirmed that the horseshoe decision rule outperforms the shrinkage rule induced by the double-exponential prior under various levels of sparsity. 
Although $\tau$ is treated as a tuning parameter that mimics $\pi$ in the theoretical treatment, in practice, $\pi$ is an unknown parameter. Several authors \citep{datta2013asymptotic, ghosh2016asymptotic, ghosh2016testing,van2016many} have shown that usual estimates of $\tau$ adapts to sparsity, a condition that also guarantees near-minimaxity in estimation. \citet{ghosh2016testing} extended the ABOS property to a wider class of global-local shrinkage priors, with conditions on the slowly varying tails of the local shrinkage prior. \citet{ghosh2016testing} prove a stronger result, namely, the testing rule under a global-local prior attains the ABOS property \textit{exactly}, when the global shrinkage parameter $\tau$ is of the same asymptotic order as the sparsity proportion $\pi$. 

\subsection{Sparse Linear Regression}\label{sec:sparse-linreg}

One of the major advantages of Lasso and other frequentist penalized methods is their theoretical optimality properties in the regression setting $\Y \sim \Nor_n(\X \btheta, \sigma^2 \I_n)$ \citep[e.g.]{buhlmann2011statistics}, whereas similar results for Bayesian methods using shrinkage priors are relatively less common. We review extant theoretical results for Bayesian sparse regression covering both point-mass mixture and continuous shrinkage priors. 

\vspace{0.1in}
\noindent \textbf{Point mass mixture priors:} Arguably the most notable contribution is due to \citet{castillo2015bayesian}, who showed that the posterior under a point-mass mixture prior contracts at the optimal rate for sparse parameter recovery and prediction, given a suitable `compatibility' condition on the design matrix $\X$ is satisfied. Such compatibility conditions also govern oracle properties for Lasso-type methods, e.g. `irrepresentability' and `mutual coherence' conditions \citep[\textit{vide} Ch. 6]{buhlmann2011statistics} and \citep{zhao2006model}. Similarly, for recovery under point-mass mixture priors, \citet{castillo2015bayesian} define three local invertibility conditions on the regression matrix: $\bar{\phi}(s)$ (uniform compatibility in sparse vectors), $\tilde{\phi}(s)$ (smallest scaled sparse singular value), and $\text{mc}(\X)$ (mutual coherence), for recovery with respect to $\ell_1$ norm, $\ell_2$ norm and $\ell_{\infty}$ norm respectively. We define the irrepresentability and mutual coherence condition below.

First, suppose the sample covariance matrix is denoted by $\hat{\Sigma} = n^{-1} \X^T \X$ and the active set $S = \{ j : \theta_j \ne 0 \}$ consists of first $s_0$ elements of $\btheta$ as in Definition \ref{def:compatibility}. One can partition the $\hat{\Sigma}$ matrix as 

\[
\hat{\Sigma} = \begin{bmatrix} 
	\hat{\Sigma}_{s_0, s_0} & \hat{\Sigma}_{s_0, p - s_0} \\
  \hat{\Sigma}_{p-s_0, s_0} & \hat{\Sigma}_{p-s_0, p - s_0}   
								\end{bmatrix},
\]

where $\hat{\Sigma}_{s_0, s_0}$ is the $s_0 \times s_0$ sub-matrix corresponding to the active variables. The strong irrepresentable condition for the variable selection consistency of Lasso is: 

\beq 
\vectornorm{ \hat{\Sigma}_{p-s_0, s_0} \hat{\Sigma}_{s_0,s_0}^{-1} sign(\btheta_S)}_{\infty} \le 1 - \eta \; \text{for positive constant vector } \eta. \label{eq:irrep}
\eeq

\citet{zhao2006model} illustrated the importance of strong irrepresentable condition on Lasso's model selection performance by showing that the probability of selecting the true sparse model is an increasing function of the irrepresentability condition number, defined as:

\beq
\eta_{\infty} = 1 - \vectornorm{ \hat{\Sigma}_{p-s_0, s_0} \hat{\Sigma}_{s_0,s_0}^{-1} sign(\btheta_S)}_{\infty} \label{eq:eta_infty}.
\eeq

The strongest of these conditions, mutual coherence ($\text{mc}(\X)$), is defined as:

\beq
\text{mc}(\X) = \max_{1 \le i \ne j \le p} \frac{\abs{\langle X_{.,i} X_{.,j} \rangle}}{\vectornorm{X_{.,i}}_2 \vectornorm{X_{.,j}}_2 }.
\label{eq:mc}
\eeq

\cite{buhlmann2011statistics} establishes the relationship between the different conditions (\textit{vide} Fig. 6.1). Clearly, these optimality results carry over to the sparse normal means problem (`sequence model') where the design matrix is identity or to regression models with an orthogonal design matrix.

\vspace{0.1in}

\noindent \textbf{Continuous shrinkage priors:} \citet{polson2010shrink} point out that the one-group priors mimic Bayesian model averaging, where one achieves better predictive performance by averaging over models supported by data, without the computational burden. Several authors \citep{polson2010shrink, polson2012local, datta2015search} have shown empirically horseshoe outperforms Lasso (as well as Bayesian model averaging) in terms of out-of-sample predictive sum-of-squares error. 


\cite{armagan2013posterior} proved posterior consistency in $p \le n$ situation for commonly used shrinkage prior including generalized double Pareto and horseshoe-type priors under simple sufficient conditions, e.g. boundedness of the eigenvalues of $\X^T\X/n$ and the number of non-zero elements $p_n = o(n/\log n)$. Under similar conditions, minimax posterior contraction rates for the Dirichlet-Laplace prior \citep{bhattacharya2014dirichlet} can be extended to the regression coefficients $\btheta$. Non-trivial extension to the high dimensional setting is still an active area. 

There are some recent developments on theoretical properties for predictive risk and variable selection properties of the horseshoe posterior under a orthogonal design matrix in $p \le n$ situation. It is worth noting that there are two slightly different approaches for specifying the horseshoe prior. First, suppose a horseshoe prior is placed directly on the regression coefficient
$\btheta$ where $p \le n$ under the model:

\begin{align*}
\y & = \X \btheta + \epsilon, \; \epsilon \sim \Nor(0, \sigma^2 I_n), \\
\theta_j & \mid \lambda_j, \tau, \sigma \sim \Nor(0, \lambda_j^2 \tau^2 \sigma^2), \\
\lambda_j & \sim f(\cdot), \tau \sim g(\cdot), \sigma \sim h(\cdot). 
\end{align*}

\citet{tang2016bayesian} proposed the half-thresholding estimator,

\[
\hat{\theta}_i^{HT} = \hat{\theta}_i^{PM} I \left(\abs{ \hat{\theta}_i^{PM}/\hat{\theta}_i^{OLS}} > \half \right),
\]

where $\hat{\theta}_i^{PM}$ and $\hat{\theta}_i^{OLS}$ are the posterior mean and the OLS solution, respectively, and showed this estimator achieves oracle property (variable selection consistency and optimal estimation) if local shrinkage priors have polynomial tails. On the other hand, \citet{bhadra2016prediction} specifies the prior on a reparametrized $\balpha$ follows (noting that $\balpha$ and $\btheta$ are one-to-one functions for a fixed design $\X$):

\begin{gather}
\y  = \X \btheta + \epsilon \stackrel{\text{reparameterize}}{\Longrightarrow} \y = \Z \balpha + \epsilon, \nonumber \\
\text{where} \; \X  = \U \D \W^T, \; (\text{singular value decomposition}), \nonumber \\
\; \Z = \U \D, \; \balpha = \W^T \btheta. \; (\text{Rank}(\D) = n). \label{eq:predrisk}
\end{gather}

Under assumption of an orthogonal design, \citet{bhadra2016prediction} investigated Stein's unbiased risk estimate for prediction, defined as $\text{SURE} = \vectornorm{\y - \tilde{\y}}^2 + 2 \sigma^2 \sum_{i=1}^{n} \dd{\tilde{y}_i}{y_i}$ for the horseshoe prior and proved that it leads to improved finite sample prediction risk, over ridge regression risk of $2n \sigma^2$. 

\begin{theorem}
Prediction risk for the purely local horseshoe regression \citep{bhadra2016prediction}. Let $\D = \I$ in \eqref{eq:predrisk} and let the global shrinkage parameter in the horseshoe regression be $\tau^2 = 1$. When true $\alpha_i = 0$, an upper bound of the component-wise risk of the purely local horseshoe regression is $1.75\sigma^2 < 2\sigma^2$. 
\end{theorem}

As pointed out before, it remains to be settled whether stronger theoretical results hold for the horseshoe or other GL priors, e.g. whether oracle properties or minimaxity results under $\ell_2$ or $\ell_1$ norm carry over to horseshoe prior in the high-dimensional set-up under compatibility or coherence conditions on the design matrix as used by \citet{buhlmann2011statistics} and \citet{castillo2015bayesian}.

\subsection{Uncertainty quantification}

Reliable uncertainty quantification is a key challenge in high-dimensional inference. While some authors \citep[e.g.,][]{chatterjee2011bootstrapping} observed that the Lasso-based estimates do not yield meaningful standard errors for the parameter estimates, \cite{castillo2015bayesian} showed poor posterior contraction for the Bayesian Lasso. These results motivate Bayesian approaches with appropriately heavy-tailed priors that produce automatic and reliable uncertainty quantification. 

\citet{chatterjee2011bootstrapping} also proposed a bootstrap-based estimator for the limiting distribution of the Lasso that attains some non-uniform consistency. Similarly, \citet{liu2013asymptotic} argue that the bootstrap could be used. But these attempts are exposed to severe super-efficiency phenomena. In contrast, \citet{zhang2014confidence} pioneered the idea to de-bias the Lasso for obtaining
an asymptotic Gaussian limiting distribution for single coordinates $\theta_i$ or other low-dimensional parameters of interest. The de-biased Lasso is of the form:
\[
\hat{\btheta}^d = \hat{\btheta}^{\text{Lasso}} + \frac{1}{n} \M \X^T (\y - \X \hat{\btheta}^{\text{Lasso}}),
\]
The matrix $M$ is constructed from the node-wise Lasso, as also advocated by \citet{van2014asymptotically} or using a convex program as proposed by \citet{javanmard2014confidence}. In both approaches, exploiting KKT conditions for the node-wise Lasso or by construction, the $\ell_\infty$ norm $\abs{\M\hat{\bSigma} - \I}_{\infty}$ is controlling the bias and $[\M\hat{\bSigma}\M]_{i,i}$ is governing the variance of the de-biased or de-sparsified Lasso.

Although one can always get confidence sets for a fixed coefficient, arguably a more specific question here is whether these credible sets (marginal credible intervals or credible $\ell_2$ balls) have both the minimax radius and the correct coverage. At the heart of these results is the impossibility theorem by \citet{li1989honest}, that says one can not construct confidence sets to be both `honest' and `adaptive' uniformly for all $\theta_0$, be it Bayesian or non-Bayesian. In particular, sparsity-adaptive credible sets can not be `honest' \citep{nickl2013confidence,li1989honest} in the sense that it is impossible to construct credible sets that have both their diameters adapt to the minimax rate for the unknown sparsity $\pi$ as well as provide nominal coverage probability over the full parameter space. 

In the context of sequence models, as \cite{van2017adaptive} point out that since the horseshoe prior achieves adaptive posterior contraction at the near-minimax rate $p_n \log(n/p_n)$ in \eqref{eq:minimax} for nearly-black objects, one needs additional conditions, e.g., excessive bias-restriction \citep{belitser2015needles} or self-similarity to ensure good coverage. 
In particular, they prove that credible balls provide uncertainty quantification up to a correct multiplicative factor, provided the sparsity proportion $\pi$ crosses the detectability threshold, $\sqrt{2 \log(n/p_n)}$. We refer the readers to Theorem 5 of \cite{van2017adaptive} for a precise statement concerning the coverage and size of the horseshoe credible sets. It appears that there is a trade-off between honesty and adaptation, and Bayesian procedures such as the horseshoe attain adaptation over honesty and de-biased methods offer honesty, often by sacrificing the optimal diameter criterion.


\section{Hyper-parameters}\label{sec:5}
Careful handling of the global shrinkage parameter $\tau$ is critical for success of the horseshoe estimator in a sparse regime as it captures the level of sparsity in the data \citep{carvalho2010horseshoe, datta2013asymptotic, van2015conditions}. However, in nearly black situations a na\"ive estimate of $\tau$ could collapse to zero, and care must be taken to prevent possible degeneracy in inference. There are two main approaches regarding choice of $\tau$: first, an empirical Bayesian approach that estimates $\tau$ from the data using a simple thresholding or maximum marginal likelihood approach (MMLE) and second, a fully Bayesian approach that specifies a hyper-prior on $\tau$.

\subsection{Marginal Likelihood} We first take a closer look at how $\tau$ affects the marginal likelihood under the horseshoe prior and the maximum marginal likelihood approach of \cite{van2017adaptive}. We can write the marginal likelihood under the horseshoe prior after marginalizing out $\theta_i$ in \eqref{eq:hs} for $\sigma^2 = 1$ from the model as:

\[
  m(y \mid \tau) = \prod_{i=1}^{n} (1+\lambda_i^2 \tau^2)^{-\half} \exp \left \{ - \frac{y_i^2}{2(1+\lambda_i^2 \tau^2)} \right \}  (1+\lambda_i^2)^{-1} d \lambda_i .
\]

\citet{tiao1965bayesian} observe that the marginal likelihood is positive at $\tau = 0$, hence the impropriety of the prior of $\tau^{-2}$ at the origin translates to the posterior. As a result, the maximum likelihood estimator of $\tau$ could potentially collapse to zero in very sparse problems \citep{polson2010shrink, datta2013asymptotic}. In \cite{van2017adaptive}, both the empirical Bayes MMLE and the full Bayes solution are restricted in the interval $[1/n,1]$ to preempt this behavior. To get the MMLE of $\tau$ using the approach of \cite{van2017adaptive}, we first calculate the marginal prior of $\theta_i$ after integrating out $\lambda_i^2$ in Equation~\eqref{eq:hs}:

\[
p_{\tau}(\theta_i) = \int_{0}^{\infty} \frac{1}{\sqrt{2 \pi}} \exp \left\{ -
  \frac{\theta_i^2}{2\lambda^2 \tau^2} \right\} \frac{1}{\lambda \tau}
  \frac{2}{\pi(1+\lambda^2)} d\lambda
  \;.
\]

The MMLE is then obtained as the maximizer of the marginal likelihood restricted to the interval $[1/n,1]$: 

\[
  \hat{\tau}_M = \argmax_{\tau \in [1/n,1]} \prod_{i=1}^{n}
  \int_{-\infty}^{\infty} \frac{1}{\sqrt{2 \pi}} \exp \left\{ - \frac{(y_i
  -\theta_i)^2}{2} \right\} p_{\tau}(\theta_i) d\theta_i
  \;.
\]
The lower bound of the maximization interval prevents a degenerate solution of $\tau$ in the sparse case. 

Handling $\tau$ is still an area of research: some papers \citep[e.g.,][]{carvalho2010horseshoe, datta2013asymptotic,piironen2017sparsity} advocate using a full Bayes approach instead of a `plug-in' maximum likelihood approach to avoid potential issues such as $\hat{\tau}$ collapsing to zero. On the other hand, \citet{van2017adaptive} note the following: 

``\textit{Piironen, Betancourt, Simpson and Vehtari close with a warning against the marginal maximum likelihood estimator. They are not the first to do so. We can only say that we have not noted problems, not in the theory and not in the simulations. We also prefer full Bayes, but the greater efficiency may weigh in 
the other direction.}" \cite[\textit{vide} Rejoinder p. 1274]{van2017adaptive}.

In practice, the MMLE approach of \cite{van2017adaptive} achieves both theoretical optimality and good numerical performance. It is computed over the interval $[1/n,1]$, which connects to the interpretation of $\tau$ as sparsity and prevents any computational issues. 

\subsection{Optimization and Cross-validation} In a recent paper, \citet{van2017adaptive} have investigated the empirical Bayes and full Bayes approach for $\tau$, and have shown that the full Bayes and the MMLE estimators achieve the near minimax rate, namely $p_n \log(n)$, under similar conditions. For the full Bayes estimator, these conditions are easily seen to satisfied by a half-Cauchy prior truncated to the interval $[1/n,1]$, which also does well in numerical experiments, both in `sparse' and `less-sparse' situations. 

The MMLE estimator of \citet{van2017adaptive} outperforms the simple thresholding estimator given by:

\[
  \hat{\tau}_s(c_1, c_2) = \max \left \{ \frac{\sum_{i=1}^{n} \1 \{ \abs{y_i} \ge
  \sqrt{c_1 \log(n) \}}}{c_2 n}, \frac{1}{n} \right\}
  \;.
\]

Rather, the MMLE estimator can detect smaller non-zero signals, even those below the threshold $\sqrt{2 \log(n)}$, such as $\theta_i = 1$ when $n = 100$. 


A third approach could be treating $\tau$ as a tuning parameter and using a $k$-fold cross-validation to select $\tau$.  As in the full Bayes and empirical Bayes approach, the cross-validated choice of $\hat{\tau}$ can also converge to zero and care should be taken to avoid such situations. Yet another approach for handling $\tau$ was proposed by \citet{piironen2016hyperprior}, who have investigated the choice of $\tau$ for a linear regression model and have suggested choosing a prior for $\tau$ by studying the prior for $m_{\text{eff}} = \sum_{i=1}^{n} (1-\kappa_i)$, the effective number of non-zero parameters. When better prediction is desired, \citet{bhadra2016prediction} suggest selecting $\tau$ by minimizing SURE, for which they provide an explicit form under the model in \eqref{eq:predrisk}. 


\section{Computation}\label{sec:horse-comp}

Over the last few years, several different implementations of the horseshoe prior for the normal means and regression models have been proposed. The MCMC based implementations usually proceed via block-updating $\btheta$, $\blambda$ and $\tau$ using either a Gibbs, parameter expansion or slice sampling strategy. The first \textsc{R} package to offer horseshoe prior for regression along with Lasso, Bayesian Lasso and ridge was the \texttt{monomvn} package by \citet{gramacy2010shrinkage}. In an unpublished technical report, \citet{scott_parameter_2010} proposed a parameter expansion strategy for the horseshoe prior and studied its effect on the autocorrelation of $\tau$. Furthermore, \citet{scott_parameter_2010} pointed out that the solution to this lies in marginalizing over the local shrinkage parameter $\lambda_j$'s. On a somewhat similar route, \citet{makalic2016high} uses a inverse-gamma scale mixture identity to construct a Gibbs sampling scheme for the horseshoe and horseshoe+ priors for linear regression as well as logistic and negative binomial regressions. 

The \texttt{horseshoe} package implements the MMLE and truncated prior approaches for handling $\tau$ proposed in \citet{van2017adaptive}. \citet{hahn_elliptical_2016} proposed an elliptical slice sampler and argue that it outperforms Gibbs strategies for higher dimensional problems both in per-sample speed and quality of samples (i.e. effective sample size). The state-of-the-art implementation for the horseshoe prior in linear regression is by \citet{bhattacharya_fast_2015} who used a Gaussian sampling alternative to the na\"ive Cholesky decomposition to reduce the computational burden from $O(p^3)$ to $O(n^2p)$. A very recent paper by \citet{james2017scalable} claims to improve this even further by implementing a block update strategy but using a random walk Metropolis--Hastings algorithm on $\log(1/\tau^2)$ for block-updating $\tau \mid \lambda$. We provide a list of all implementations known to us on Table \ref{tab:hs-imp}. 

Bayesian methods using MCMC are sequential in nature. The methods are typically computation intensive, but one is able to perform probabilistic uncertainty quantification. However, sparse Bayesian methods including the horseshoe regression can be computed for $p \approx 10^6$, using parallel architecture of the latent variable representation to be able to retain the  fully Bayesian nature via MCMC sampling. \cite{terenin_gpu-accelerated_2016} implement a horseshoe-probit regression using GPU that takes $\approx 2$ minutes for calculations involving a design matrix $\X$ of dimensions $10^6 \times 10^3$. If only point estimates are desired, of course Bayesian posterior modes can be computed as fast as penalized likelihood estimates \citep{bhadra2017horseshoe}.

\begin{table}[htbp]
  \centering
  \caption{Implementations of the horseshoe and other shrinkage priors}
  \footnotesize{
    \begin{tabular}{|c|c|}
    \hline
    Implementation (Package/URL) & Authors \bigstrut\\
    \hline
    \textsc{R} package: \href{https://cran.r-project.org/web/packages/monomvn/index.html}{\texttt{monomvn}} & \citet{gramacy2010shrinkage} \bigstrut[t]\\
     \textsc{R} code in paper & \citet{scott_parameter_2010} \\
    \textsc{R} package: \href{https://cran.r-project.org/web/packages/horseshoe/index.html}{\texttt{horseshoe}} & \citet{pas_horseshoe:_2016} \\
    \textsc{R} package: \href{https://cran.r-project.org/web/packages/fastHorseshoe/index.html}{\texttt{fastHorseshoe}} & \citet{hahn_elliptical_2016} \\
    \href{https://github.com/antik015/Fast-Sampling-of-Gaussian-Posteriors}{\textsc{Matlab} code} & \citet{bhattacharya_fast_2015} \\
    GPU accelerated Gibbs sampling & \citet{terenin_gpu-accelerated_2016} \\
    \href{https://cran.r-project.org/web/packages/bayesreg/index.html}{\texttt{bayesreg}} + \textsc{Matlab} code in paper & \citet{makalic2016high} \\
     \href{https://github.com/jamesjohndrow/horseshoe_jo}{\textsc{Matlab} code} & \citet{james2017scalable} \bigstrut[b]\\ 
		R package:\href{https://cran.r-project.org/web/packages/bayeslm/index.html}{\texttt{bayeslm}} & \citet{hahn2017efficient} \\
    \hline
    \end{tabular}%
    }
  \label{tab:hs-imp}%
\end{table}%


\section{Simulation Experiments}\label{sec:simulation}

\subsection{Effect of Correlated Predictors}

As we discussed in \S \ref{sec:sparse-linreg}, Lasso as well as Bayesian spike-and-slab priors can recover regression parameters under strong assumptions on the design matrix such as `irrepresentability' or `mutual coherence'. As \citet{van2017adaptive} point out, such conditions are expected to be necessary for optimal recovery as in the context of spike-and-slab prior \citep{castillo2015bayesian}. 

For this simulation study, we follow the set-up in \citet{zhao2006model} closely. Let $S = \{j : \theta_{j0} \ne 0 \}$ be the active set of predictors, and let $s_0 = \abs{S}$. We simulate data with $n = 100, p = 60$ and $s_0 = 7$ with the sparse coefficient vector $\btheta_S^* = (7, 5, 5, 4, 4, 3, 3)^T$ . The error variance $\sigma^2$ was set to $0.1$ to obey the asymptotic properties of the Lasso.

We first draw the covariance matrix $\Sigma$ from $\text{Wishart}(p, I_p)$ and then generate design matrix $\X$ from $\Nor(0, \Sigma)$. \citet{zhao2006model} showed that the Strong Irrepresentability Condition \eqref{eq:irrep} may not hold for such a design matrix. We generate $100$ such design matrices to obtain a range of different $\eta_{\infty}$ values. In our simulation studies the $\eta_{\infty}$ values in \eqref{eq:eta_infty} for the 100 simulated designs were between $[-0.86, 0.38]$. To see how the irrepresentability condition affects probability of selecting the correct model, $100$ simulations were conducted for each design matrix. We compare four different methods: two penalized likelihood methods: Lasso, SCAD (Smoothly Clipped Absolute Deviation) \citep{fan2001variable}, and two Bayesian methods: horseshoe and Dirichlet--Laplace \citep{bhattacharya2014dirichlet} in terms of percentage of these methods selecting the correct model. For model selection, we use the credible intervals for the horseshoe prior and $k$-means clustering for the Dirichlet--Laplace prior, following the simulation study in \citet{bhattacharya2014dirichlet}. 

Like \cite{zhao2006model}, we expect the Lasso to select the true model with a high probability when $\eta_{\infty} >0$ and poorly when $\eta_{\infty} < 0$, with the sharpest ascent around the origin. We also calculated the mutual coherence \eqref{eq:mc} number for the same design matrices to see the effect on these two methods. The $\text{mc}(\X)$ numbers were between $[0.21, 0.54]$. 

\begin{figure}[ht!]%
\centering
\includegraphics[height = 3in, width=\columnwidth]{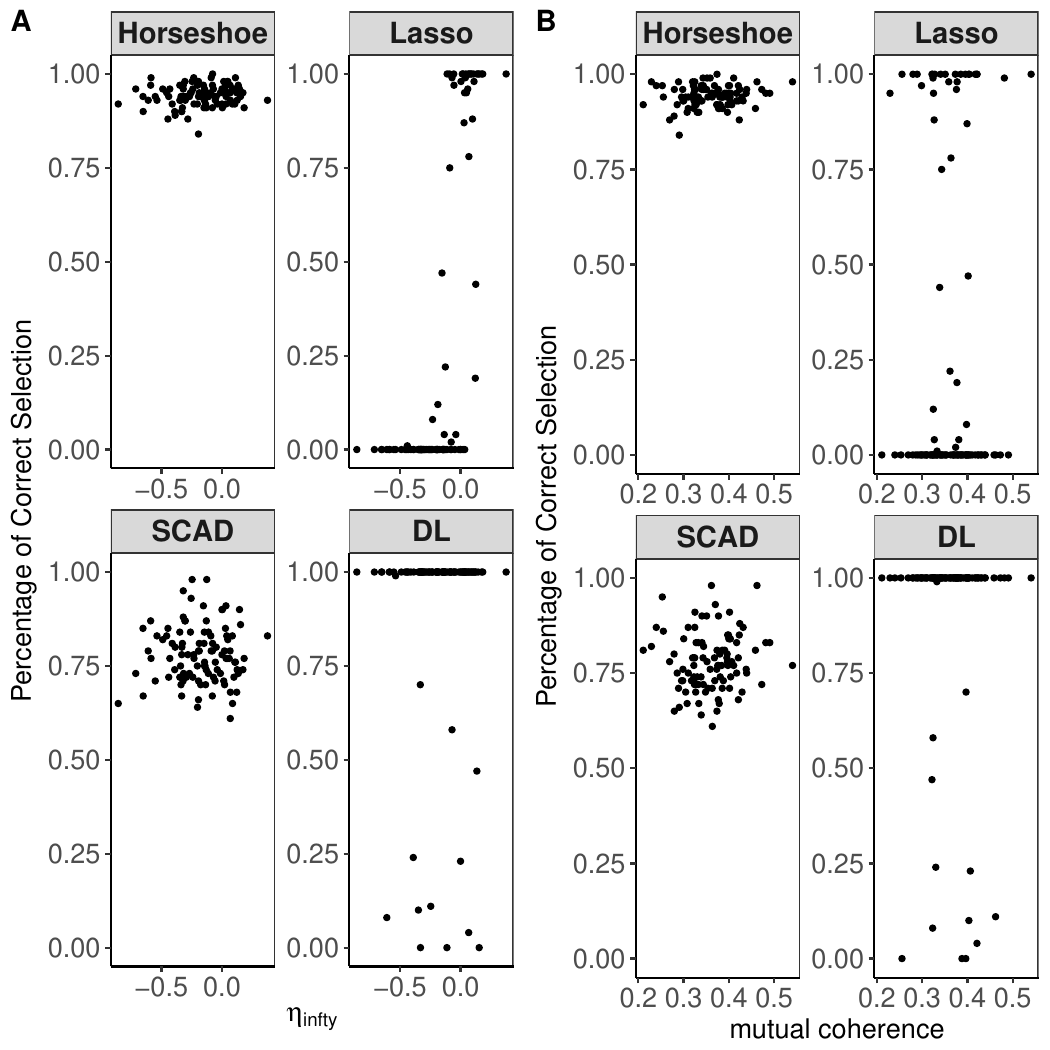} 
\caption{Effect of Strong Irrepresentability Condition $\eta_{\infty}$ (Panel A)  and Mutual Coherence or 
maximum column correlation (Panel B) on the percentage of selecting the correct model
by Lasso and SCAD penalties as well as the horseshoe and Dirichlet--Laplace (DL) priors.}%
\label{fig:irrep-mc}%
\end{figure}

Figure \ref{fig:irrep-mc} shows the percentage of correctly selected model as a function of the irrepresentable condition number, $\eta_{\infty}$ and mutual coherence  for the four candidates: Lasso, SCAD, horseshoe and Dirichlet--Laplace. For this simulation experiment, Lasso's model selection performance is dependent on the irrepresentability condition, deteriorating with decreasing $\eta_{\infty}$. Surprisingly, the effect is weaker for SCAD as well as both the horseshoe and the Dirichlet--Laplace priors. 

While the horseshoe almost always recovers the true sparse $\btheta$ vector irrespective of $\eta_{\infty}$, SCAD exhibits a high percentage (mean = $0.75$, range = $[0.61, 0.98]$). Since we have calculated mutual coherence for the same design matrices, in this set-up it does not affect the horseshoe prior's variable selection, and its effect shows no clear pattern on any other candidates, apart from the Lasso. 

\subsection{Binary Response: Logistic Regression}

We compare the performance of the horseshoe prior and Lasso for logistic regression for varying degree of dependence between the columns of a design matrix. We generated $n = 100$ binary observations for the standard logistic regression. The true parameter $\btheta^* \in \mathbb{R}^p$ where $p = 32$, $\btheta^*$ is sparse and has 5 non-zero elements $(7, 4, 2, 1, 1)$, and $\sigma^2$ was set to $0.1$. We set the covariance matrix as $\Sigma_{ij} = \rho^{\abs{i-j}}$ and then generate design matrix $\X$ from $\Nor(0, \Sigma)$ for $20$ different values of $\rho \in [0.1, 0.9]$. Since the original horseshoe prior was not designed to handle the logistic likelihood, we use the Gaussian approximation method by \cite{piironen2017sparsity}, where they use a second-order Taylor expansion for the log posterior distribution. \cite{piironen2017sparsity} also propose the regularized horseshoe prior where one introduces an additional slab width $c$ to allow for shrinkage even on the extreme tails. Following the recommendations of \cite{piironen2017sparsity}, we use the regularized horseshoe prior with a hyper-prior $c \sim \text{Inv-Gamma}(2,8)$ that corresponds to a $\text{Student-t}(0, 2^2)$ slab. We use $1,000$ posterior draws per chain with the NUTS algorithm in Stan. For Lasso, we use the \texttt{glmnet} package in R with a $10$-folds cross-validation. 

To compare the two methods for classification and predictive accuracy, we train the models on 80\% of the data, with the remaining as test set and average the results over $50$ random splits. We measure classification accuracy by the number of misclassified response $y_i$'s in test data. For predictive accuracy, we compare the mean log predictive density (MLPD) proposed in \cite{gelman2014understanding} as the mean of the computed log pointwise predictive density, defined as follows. 

Let $\btheta^s$; $s = 1, \ldots, S$ be the posterior draws from $p(\btheta \mid \y)$, and $\y_j$, $j = 1, \ldots, m$ be the $j^{th}$ test data, then MLPD is:

\begin{equation}
\text{MLPD} = \frac{1}{m} \sum_{j=1}^{m} \log \left( \frac{1}{S} \sum_{s=1}^{S} p(\y_j \mid \btheta^s) \right).
\label{eq:mlpd}
\end{equation}

\begin{figure}[t!]%
\begin{subfigure}[t]{0.45\linewidth}%
\includegraphics[height=2in,width=\columnwidth]{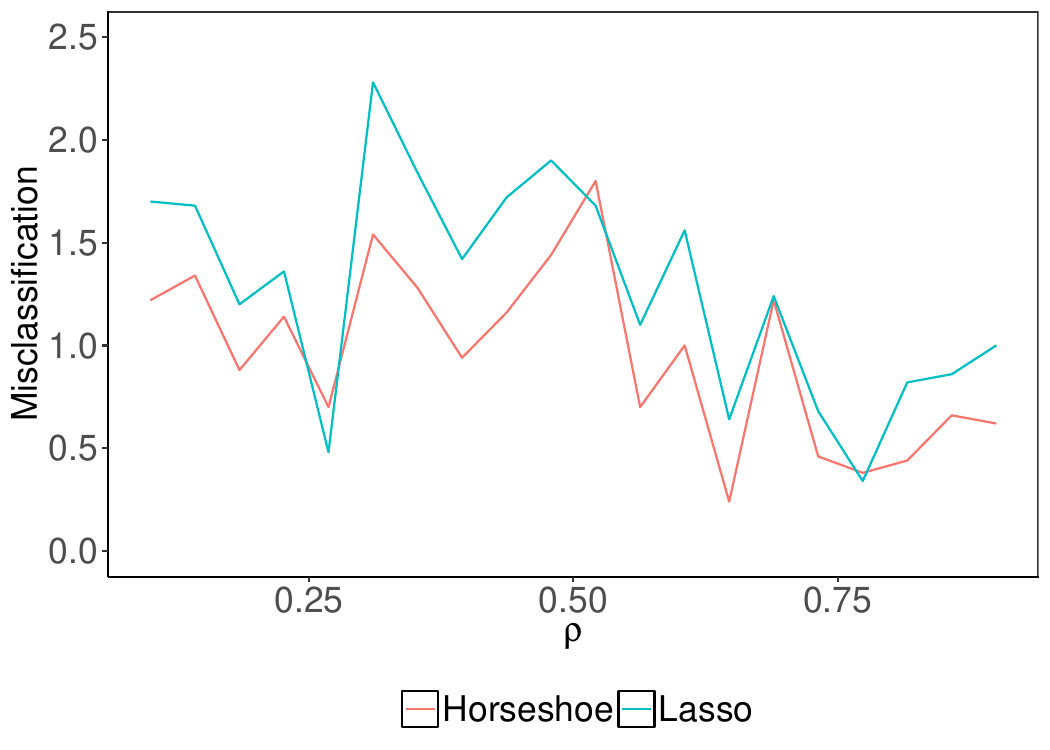}%
\label{fig:4a}%
\vspace{-0.5cm}
\begin{center}
(a)
\end{center}
\end{subfigure}
\begin{subfigure}[t]{0.45\linewidth}%
\includegraphics[height=2in,width=\columnwidth]{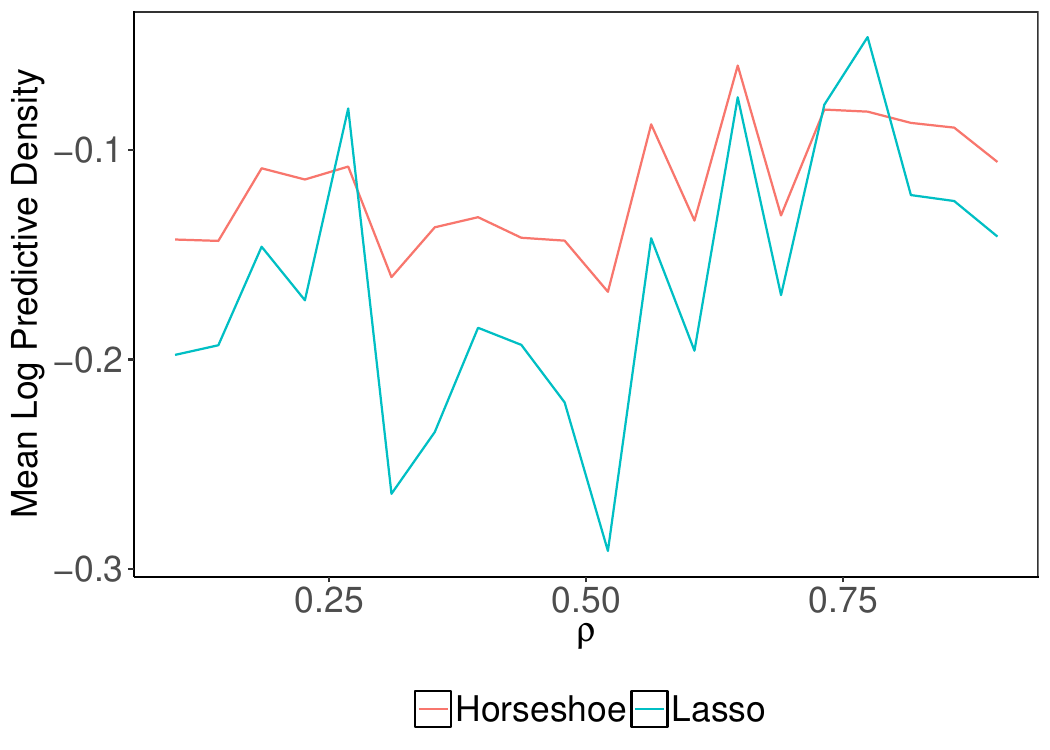}%
\label{fig:4b}%
\vspace{-0.5cm}
\begin{center}
(b)
\end{center}
\end{subfigure}
\caption{(a) Number of misclassified test data points and (b) mean log predictive
density in \eqref{eq:mlpd} by the horseshoe and Lasso across different values of correlation
$\rho$, where a higher value of $\rho$ represents higher dependence between the columns of $\X$.}%
\label{fig:logistic}%
\end{figure}

Figure \ref{fig:logistic}(a) shows the average number of misclassified observations by horseshoe is a little lower than Lasso for all but two values of $\rho$. For the same values of $\rho$, Fig. \ref{fig:logistic}(b) shows that the predictive accuracy under the horseshoe prior is a little better than the Lasso. We direct the readers to \cite{piironen2017sparsity} for a thorough comparison between the different variants of the horseshoe prior with Lasso for a few real data set as well as a synthetic data-set with a separable predictor.

\section{Further Developments}\label{sec:app-ext}


\subsection{Further Developments on Lasso} Since the inception of Lasso as a regularization method for linear regression in 1996, a great deal of extensions and applications have been proposed in the literature. The combined effect of convex penalty and sparsity of the final solution lead to huge computational gains by using powerful convex optimization methods on problems of massive dimensions. The coordinate descent approach \citep{friedman_pathwise_2007, friedman2010regularization} is one particularly promising approach, that works by applying soft-threshold to the least-squares solution obtained on partial residuals, one at a time. The coordinate descent approach is flexible and easy and can be proved to converge to the solution as long as the log-likelihood and penalty are convex \citep{tseng2001convergence}, paving the way for wide applicability of $\ell_1$ penalty in generalized linear models (GLM). The popular R package \texttt{glmnet} provides a nice and easy interface for applying Lasso and elastic-net penalty for a general sparse GLM.


\subsection{Further Developments on Horseshoe}

As discussed in Section~\ref{sec:one-gp}, the horseshoe prior belongs to a wider class of global-local shrinkage priors \citep{polson2010shrink} that are characterized by a local shrinkage parameter for recovering large signals and a global shrinkage parameter for adapting to overall sparsity. The class of global-local priors, although differing in their specific goals and design, exhibit some common features: heavy tails for tail-robustness and appreciable mass near zero for sparsity, leading to shared optimality properties. 

Although the original horseshoe prior was developed for signal recovery with sparse Gaussian means, the idea of directly modeling the posterior inclusion probabilities and the use of normal-scale mixtures to facilitate sparsity is very flexible and can be easily generalized to a wider class of problems. \citet{bhadra2015default} show that the horseshoe prior is a good candidate as a default prior for low-dimensional, possibly non-linear functionals of high-dimensional parameter and can resolve long-standing marginalization paradoxes for such problems. \citet{bhadra2016prediction} show how to use global-local priors for prediction and provide theoretical and numerical evidence that it performs better than a variety of competitors including Lasso, ridge, PCR and sparse PLS. 

Moving beyond Gaussianity, \citet{datta2016bayesian} re-discovered the Gauss hypergeometric prior for flexible shrinkage needed for quasi-sparse count data, with a tighter control on false discoveries.
\citet{piironen2016hyperprior} used a Gaussian approximation using a second-order Taylor expansion for the log-likelihood to apply the horseshoe prior in generalized linear models. \citet{wang2013class} proposed a shrinkage prior based on a scale mixture of uniform for covariance matrix estimation. \citet{peltola2014hierarchical} applied the horseshoe prior for Bayesian linear survival regression for selecting covariates with highest predictive values. A sample of the many applications of the horseshoe prior is given in Table \ref{tab:hs-apps}. Given the explosive growth of methodology in
this area, we conjecture that the horseshoe prior would be regarded as a key tool for sparse signal recovery and as a default prior for objective Bayesian inference in many important problems.

\begin{table}[!t]
  \centering
  \caption{Applications of the horseshoe prior}
  \footnotesize{
    \begin{tabular}{|p{3in}|p{1.5in}|}
    \hline
    Application  & Authors \bigstrut\\
    \hline
    \textit{Fadeout} method for mean-field variational inference under non-centered parameterizations and stochastic variational inference for undirected graphical model.  & \citet{ingraham_bayesian_2016} \bigstrut[t]\\
    \hline
    Linear regression for Causal inference and Instrumental variable models  & \citet{hahn_shrinkage_2014, hahn_elliptical_2016} \\
    \hline
    Multiclass prediction using DOLDA (Diagonally orthant Latent Dirichlet Allocation)  & \citet{magnusson_dolda_2016} \\
    \hline
    Mendelian Randomization to detect causal effects of interest & \citet{berzuini_mendelian_2016} \\
    \hline 
    Locally adaptive nonparametric curve fitting with shrinkage prior Markov random field (SPMRF) & \citet{faulkner_bayesian_2015} \\
    \hline
    Quasi-Sparse Count Data & \citet{datta2016bayesian} \\
    \hline
    Variable Selection under the projection predictive framework  & \citet{piironen_projection_2015} \bigstrut[b]\\
    \hline
  Dynamic shrinkage Process (dynamic linear model and trend filtering) & \citet{kowal2017dynamic} \\
   \hline 
   Logistic regression with horseshoe prior & \citet{piironen2017sparsity, wei2017bayesian} \\
  \hline  
  Tree ensembles with rule structured horseshoe regularization & \citet{nalenz2017tree} \\
 \hline 
Bayesian compression for deep learning & \citet{louizos2017bayesian} \\
 \hline 
Precision matrix estimation & \citet{li2017graphical} \bigstrut[b]\\
\hline
    \end{tabular}%
    }
  \label{tab:hs-apps}%
\end{table}%


\section{Discussion}\label{sec:9}

Sparsity can be achieved with Lasso and horseshoe regularization, a member of the class of global-local shrinkage priors. The horseshoe prior offers better computational efficiency than the Bayesian two-group priors, while still mimicking the inference and it outperforms the estimator based on Laplace prior, the Bayesian dual of Lasso. The intuitive reasons for better performance by the horseshoe prior are its heavy tails and probability spike at zero, which make it adaptive to sparsity and robust to large signals. A number of computing strategies have been proposed for both the Lasso and the horseshoe prior, based on variants of coordinate descent and MCMC respectively.  We have outlined the distinct algorithmic implementations in \S \ref{sec:horse-comp} and Table~\ref{tab:hs-imp}. Since the goal of Lasso-based estimator is to produce a point estimate, rather than samples from the full posterior distribution of the underlying parameter, Lasso-based methods are typically faster than the horseshoe and related shrinkage priors.

The lack of speed can be overcome easily by employing a strategy based on expectation-maximization or proximal algorithm, which is often faster than the Lasso or other penalty based methods, for example the EM algorithm proposed in \S 4 of \citet{bhadra2017horseshoe} is orders of magnitude faster than the non-convex SCAD or MCP \citep[\textit{vide} Table 1]{bhadra2017horseshoe}. Another fruitful strategy is to employ proximal algorithms similar to expectation-maximization \citep{polson2015proximal}. These algorithms can be specifically designed to achieve better estimation and prediction error compared to Lasso \citep{bhadra2017horseshoe} by using clever decompositions of the objective function and some convenient properties (e.g. strong convexity) of the resulting parts. As discussed before, an active area of research is designing algorithms to handle Bayesian shrinkage in big data problems, e.g. using GPU-accelerated computing \citep{terenin_gpu-accelerated_2016}. 

We have discussed the theoretical optimality properties for both Lasso and horseshoe estimators. The optimality properties of Lasso in regression are well-known and they depend on the `neighborhood stability' or `irrepresentability' condition \eqref{eq:irrep} and the `beta-min' condition. Similarly, adaptive posterior concentration for horseshoe depends on `excessive bias restriction', a condition analogous to the `beta-min' condition. Although horseshoe regression has not been studied to the same depth as penalized regression, it is expected that optimality will depend on conditions that guarantee against ill-posed design matrix and separability of signal and noise parameters. For the sequence model, the horseshoe posterior mean enjoys near-minimaxity in estimation, and the induced decision rule achieves asymptotic Bayes optimality for multiple testing as discussed in Section~\ref{sec:stat-prop}. 

The horseshoe estimator of the sampling density converges to the the true sampling density $p(y \mid \theta_0)$ at a super-efficient rate at $\theta_0 = 0$, compared to any Bayes estimator with a bounded prior density at the origin \citep[\textit{vide} Theorem 4]{carvalho2010horseshoe}. The rate of convergence of the Ces\'aro-average Bayes risk at $\theta_0 = 0$ for horseshoe is $O(n^{-1}(\log n - b \log \log n))$. This is called the `Kullback--Leibler super-efficiency' in true density recovery for the horseshoe estimator. The horseshoe priors are also good default priors for many-to-one functionals as shown in \citet{bhadra2015default}, but a thorough study of horseshoe prior for default Bayes problems is still an unexplored area. We end the current article with a few other possible directions for future investigations.
\begin{enumerate}
   \item The square-root Lasso \citep{belloni2011square} or scaled Lasso \citep{sun2012scaled} improves over the Lasso by making the inference ambivalent towards $\sigma$, while making the estimator scale-invariant. It might be interesting to study the effect of marginalizing the global parameters such as $\tau$ and $\sigma$ on inference from shrinkage priors. Our preliminary investigation suggests that scaling the prior on $\tau$ by $\sigma$ or marginalizing out $\sigma$ improves the robustness of the shrinkage priors.
	\item One promising area is to extend the inferential capacity for the exponential family, and whether or not the optimality properties carry over to the non-Gaussian cases. Some early research on this is \citet{datta2016bayesian} and \citet{wei2017bayesian}. 
	\item Another interesting direction could include structured sparsity under the horseshoe prior, such as grouped variable selection and Gaussian graphical models, as explored in \cite{li2017graphical}. 
\end{enumerate}

\begin{appendix}





\section{Two-groups Model}\label{sec:2gp}

The two-groups model is a natural hierarchical Bayesian model for the sparse
signal-recovery problem.  The two-groups solution to the signal detection
problem is as follows:
\begin{enumerate}
\item Assume each $\theta_i$ is non-zero with some common prior probability $(1 - \pi)$, and that the nonzero $\theta_i$ come from a common density $\Nor(0,\psi^2)$. 
\item Calculate the posterior probabilities that each $y_i$ comes from $\Nor(0,\psi^2)$. 
\end{enumerate}
The most important aspect of this model is that it automatically adjusts for multiplicity without any ad-hoc regularization, i.e. it lets the data choose $\pi$ and then carry out the tests on the basis of the posterior inclusion probabilities $\omega_i = P(\theta_i \neq 0 \mid y_i)$. Formally, in a two-groups model $\theta_i$'s are modeled as

\begin{equation}
\theta_i \mid \pi, \psi = (1-\pi)\delta_{0} + \pi \Nor (0, \psi^2), \label{spikeslab}
\end{equation}

where $\delta_{0}$ denotes a point mass at zero and the parameter $\psi^2>0$ is
the non-centrality parameter that determines the separation between the two
groups. Under these assumptions, the marginal distribution of $(y_i \mid \pi, \psi)$
is given by:

\begin{equation}
y_i \mid \pi, \psi \sim  (1-\pi) \Nor(0, 1) + \pi \Nor(0, 1+\psi^2). \label{twogroups}
\end{equation}

From \eqref{twogroups}, we see that the two-groups model leads to a sparse
estimate, i.e., it puts exact zeros in the model. 

\section{Proof of Equation (3.5)}\label{sec:proof}

Assume $\sigma^2 = 1$ without loss of generality. The hierarchical model for horseshoe prior is $y_i \sim \Nor(\theta_i,1)$ and $\theta_i \sim \Nor(0, \lambda_i^2 \tau^2)$. Using Bayes' rule, posterior density of $\theta_i$ is Gaussian with mean $(1-\kappa_i)y_i$ where $\kappa_i = 1 / (1 + \lambda_i^2 \tau^2)$. It follows from Fubini's theorem:
\[
E(\theta_i \mid y_i) = \int_0^1 (1-\kappa_i)y_i p(\kappa_i \mid y_i) d \kappa_i = \{1 - E(\kappa_i \mid y_i)\} y_i.
\]

\section{Shrinkage Profiles}\label{sec:shrink}
We compare the shrinkage functions for Lasso, ridge, and the horseshoe estimator with that of the post-lava estimator \citep{chernozhukov2017lava}. The shrinkage functions for these methods are given below:

\begin{align}
d_{\text{lasso}}(z) & = \argmin_{\theta \in \mathbb{R}} \{ (z-\theta)^2 + \lambda_l \abs{\theta} \} = (\abs{z} - \lambda_l/2)_{+} sgn(z), \\
d_{\text{ridge}}(z) & = \argmin_{\theta \in \mathbb{R}} \{ (z-\theta)^2 + \lambda_r \theta^2 \}= (1 + \lambda_r)^{-1} z, \\
d_{\text{post-lava}}(z) & = \begin{cases} 
                                         z & \quad \abs{z} > \lambda_1/2k, \\
                                         (1-k)z & \quad \abs{z} \le \lambda_1/2k,
                                        \end{cases}
\; \text{where} \; k = \lambda_2/ (1 + \lambda_2). \\
d_{\text{horseshoe}}(z) & = z \left( 1 - \frac{2 \Phi_1(1/2,1,5/2,z^2/2,1-1/\tau^2)}{3\Phi_1(1/2,1,3/2, z^2/2, 1 - 1/\tau^2)} \right).
\end{align}

Figure \ref{fig:score} shows the post-lava and the horseshoe shrinkage function along with Lasso and ridge shrinkage functions for $z > 0$. Although a theoretical analysis is beyond the scope of the current article, we can see the similarities between the lava and horseshoe shrinkage. They both shrink aggressively for small values of $z$ and provide robustness for large signals $z$, as the shrinkage function becomes closer to the $45^\circ$ line.
\end{appendix}

\section*{Acknowledgements}
We thank the AE and two anonymous referees for constructive suggestions. Bhadra and Polson are partially supported by Grant No. DMS-1613063 by the US National Science Foundation.

\bibliographystyle{authordate1}
\bibliography{hs-review}

\begin{thebibliography}{}

\bibitem[\protect\citename{Andrews \& Mallows, }1974]{andrews1974scale}
Andrews, David~F, \& Mallows, Colin~L. 1974.
\newblock Scale mixtures of normal distributions.
\newblock {\em Journal of the Royal Statistical Society. Series B
  (Methodological)}, {\bf 36}, 99--102.

\bibitem[\protect\citename{Armagan {\em et~al.}, }2011]{armagan2011generalized}
Armagan, Artin, Clyde, Merlise, \& Dunson, David~B. 2011.
\newblock Generalized Beta Mixtures of {{Gaussians}}.
\newblock {\em Pages  523--531 of:} {\em Advances in {{Neural Information
  Processing Systems}}}.

\bibitem[\protect\citename{Armagan {\em et~al.},
  }2013a]{armagan2013generalized}
Armagan, Artin, Dunson, David~B, \& Lee, Jaeyong. 2013a.
\newblock Generalized Double {{Pareto}} Shrinkage.
\newblock {\em Statistica Sinica}, {\bf 23}(1), 119--143.

\bibitem[\protect\citename{Armagan {\em et~al.}, }2013b]{armagan2013posterior}
Armagan, Artin, Dunson, David~B, Lee, Jaeyong, Bajwa, Waheed~U, \& Strawn,
  Nate. 2013b.
\newblock Posterior consistency in linear models under shrinkage priors.
\newblock {\em Biometrika}, {\bf 100}(4), 1011--1018.

\bibitem[\protect\citename{Bai \& Ghosh, }2017]{bai2017inverse}
Bai, Ray, \& Ghosh, Malay. 2017.
\newblock The Inverse Gamma-Gamma Prior for Optimal Posterior Contraction and
  Multiple Hypothesis Testing.
\newblock {\em arXiv preprint arXiv:1710.04369}.

\bibitem[\protect\citename{Belitser \& Nurushev, }2015]{belitser2015needles}
Belitser, Eduard, \& Nurushev, Nurzhan. 2015.
\newblock Needles and straw in a haystack: robust confidence for possibly
  sparse sequences.
\newblock {\em arXiv preprint arXiv:1511.01803}.

\bibitem[\protect\citename{Belloni {\em et~al.}, }2011]{belloni2011square}
Belloni, Alexandre, Chernozhukov, Victor, \& Wang, Lie. 2011.
\newblock Square-Root Lasso: Pivotal Recovery of Sparse Signals via Conic
  Programming.
\newblock {\em Biometrika}, {\bf 98}(4), 791--806.

\bibitem[\protect\citename{Berzuini {\em et~al.},
  }2016]{berzuini_mendelian_2016}
Berzuini, Carlo, Guo, Hui, Burgess, Stephen, \& Bernardinelli, Luisa. 2016.
\newblock Mendelian {Randomization} with {Poor} {Instruments}: a {Bayesian}
  {Approach}.
\newblock {\em arXiv:1608.02990 [math, stat]}, Aug.
\newblock arXiv: 1608.02990.

\bibitem[\protect\citename{Bhadra {\em et~al.}, }2016a]{bhadra2015default}
Bhadra, Anindya, Datta, Jyotishka, Polson, Nicholas~G, \& Willard, Brandon.
  2016a.
\newblock Default {B}ayesian analysis with global-local shrinkage priors.
\newblock {\em Biometrika}, {\bf 103}(4), 955--969.

\bibitem[\protect\citename{Bhadra {\em et~al.}, }2016b]{bhadra2016global}
Bhadra, Anindya, Datta, Jyotishka, Polson, Nicholas~G, \& Willard, Brandon.
  2016b.
\newblock Global-{{Local Mixtures}}.
\newblock {\em arXiv preprint arXiv:1604.07487}.

\bibitem[\protect\citename{Bhadra {\em et~al.}, }2016c]{bhadra2016prediction}
Bhadra, Anindya, Datta, Jyotishka, Li, Yunfan, Polson, Nicholas~G, \& Willard,
  Brandon. 2016c.
\newblock Prediction Risk for the Horseshoe Regression.
\newblock {\em arXiv preprint arXiv:1605.04796}.

\bibitem[\protect\citename{Bhadra {\em et~al.}, }2017a]{bhadra2015horseshoe+}
Bhadra, Anindya, Datta, Jyotishka, Polson, Nicholas~G, \& Willard, Brandon.
  2017a.
\newblock The {{Horseshoe}}+ {{Estimator}} of {{Ultra}}-{{Sparse Signals}}.
\newblock {\em Bayesian Analysis}, {\bf 12}(4), 1105--1131.

\bibitem[\protect\citename{Bhadra {\em et~al.}, }2017b]{bhadra2017horseshoe}
Bhadra, Anindya, Datta, Jyotishka, Polson, Nicholas~G, \& Willard, Brandon.
  2017b.
\newblock Horseshoe {{Regularization}} for {{Feature Subset Selection}}.
\newblock {\em arXiv preprint arXiv:1702.07400}.

\bibitem[\protect\citename{Bhattacharya {\em et~al.},
  }2015]{bhattacharya2014dirichlet}
Bhattacharya, Anirban, Pati, Debdeep, Pillai, Natesh~S., \& Dunson, David~B.
  2015.
\newblock Dirichlet-{{Laplace}} Priors for Optimal Shrinkage.
\newblock {\em Journal of the American Statistical Association}, {\bf 110},
  1479--1490.

\bibitem[\protect\citename{Bhattacharya {\em et~al.},
  }2016]{bhattacharya_fast_2015}
Bhattacharya, Anirban, Chakraborty, Antik, \& Mallick, Bani~K. 2016.
\newblock Fast sampling with Gaussian scale mixture priors in high-dimensional
  regression.
\newblock {\em Biometrika},  asw042.

\bibitem[\protect\citename{Bien {\em et~al.}, }2013]{bien_lasso_2013}
Bien, Jacob, Taylor, Jonathan, \& Tibshirani, Robert. 2013.
\newblock A Lasso for Hierarchical Interactions.
\newblock {\em Annals of Statistics}, {\bf 41}(3), 1111--1141.

\bibitem[\protect\citename{Bogdan {\em et~al.}, }2011]{bogdan2011asymptotic}
Bogdan, Ma{\l}gorzata, Chakrabarti, Arijit, Frommlet, Florian, \& Ghosh,
  Jayanta~K. 2011.
\newblock Asymptotic {{Bayes}}-Optimality under Sparsity of Some Multiple
  Testing Procedures.
\newblock {\em The Annals of Statistics}, {\bf 39}(3), 1551--1579.

\bibitem[\protect\citename{B{\"u}hlmann \& {van de Geer},
  }2011]{buhlmann2011statistics}
B{\"u}hlmann, Peter, \& {van de Geer}, Sara. 2011.
\newblock {\em Statistics for High-Dimensional Data}.
\newblock {Springer-Verlag Berlin Heidelberg}.

\bibitem[\protect\citename{Candes \& Tao, }2007]{candes2007dantzig}
Candes, Emmanuel, \& Tao, Terence. 2007.
\newblock The {{Dantzig}} Selector: {{Statistical}} Estimation When p Is Much
  Larger than N.
\newblock {\em The Annals of Statistics},  2313--2351.

\bibitem[\protect\citename{Candes, }2008]{candes2008restricted}
Candes, Emmanuel~J. 2008.
\newblock The Restricted Isometry Property and Its Implications for Compressed
  Sensing.
\newblock {\em Comptes Rendus Mathematique}, {\bf 346}(9-10), 589--592.

\bibitem[\protect\citename{Cand{\`e}s \& Tao, }2010]{candes2010power}
Cand{\`e}s, Emmanuel~J, \& Tao, Terence. 2010.
\newblock The Power of Convex Relaxation: {{Near}}-Optimal Matrix Completion.
\newblock {\em IEEE Transactions on Information Theory}, {\bf 56}(5),
  2053--2080.

\bibitem[\protect\citename{Carvalho {\em et~al.}, }2009]{carvalho2009handling}
Carvalho, C.~M., Polson, N.~G., \& Scott, J.~G. 2009.
\newblock Handling Sparsity via the Horseshoe.
\newblock {\em Journal of Machine Learning Research W\&CP}, {\bf 5}, 73--80.

\bibitem[\protect\citename{Carvalho {\em et~al.}, }2010]{carvalho2010horseshoe}
Carvalho, Carlos~M, Polson, Nicholas~G, \& Scott, James~G. 2010.
\newblock The Horseshoe Estimator for Sparse Signals.
\newblock {\em Biometrika}, {\bf 97}, 465--480.

\bibitem[\protect\citename{Castillo \& {van der Vaart},
  }2012]{castillo2012needles}
Castillo, Isma{\"e}l, \& {van der Vaart}, Aad. 2012.
\newblock Needles and Straw in a Haystack: {{Posterior}} Concentration for
  Possibly Sparse Sequences.
\newblock {\em The Annals of Statistics}, {\bf 40}(4), 2069--2101.

\bibitem[\protect\citename{Castillo {\em et~al.}, }2015]{castillo2015bayesian}
Castillo, Ismail, Schmidt-Hieber, Johannes, \& {van der Vaart}, Aad. 2015.
\newblock Bayesian Linear Regression with Sparse Priors.
\newblock {\em The Annals of Statistics}, {\bf 43}(5), 1986--2018.

\bibitem[\protect\citename{Chatterjee \& Lahiri,
  }2011]{chatterjee2011bootstrapping}
Chatterjee, Arindam, \& Lahiri, Soumendra~Nath. 2011.
\newblock Bootstrapping lasso estimators.
\newblock {\em Journal of the American Statistical Association}, {\bf
  106}(494), 608--625.

\bibitem[\protect\citename{Chernozhukov {\em et~al.},
  }2017]{chernozhukov2017lava}
Chernozhukov, Victor, Hansen, Christian, Liao, Yuan, {\em et~al.} 2017.
\newblock A lava attack on the recovery of sums of dense and sparse signals.
\newblock {\em The Annals of Statistics}, {\bf 45}(1), 39--76.

\bibitem[\protect\citename{Cutillo {\em et~al.}, }2008]{cutillo2008larger}
Cutillo, Luisa, Jung, Yoon~Young, Ruggeri, Fabrizio, \& Vidakovic, Brani. 2008.
\newblock Larger posterior mode wavelet thresholding and applications.
\newblock {\em Journal of Statistical Planning and Inference}, {\bf 138}(12),
  3758--3773.

\bibitem[\protect\citename{Datta \& Dunson, }2016]{datta2016bayesian}
Datta, Jyotishka, \& Dunson, David~B. 2016.
\newblock Bayesian Inference on Quasi-Sparse Count Data.
\newblock {\em Biometrika}, {\bf 103}(4), 971--983.

\bibitem[\protect\citename{Datta \& Ghosh, }2013]{datta2013asymptotic}
Datta, Jyotishka, \& Ghosh, Jayanta~K. 2013.
\newblock Asymptotic Properties of {{Bayes}} Risk for the Horseshoe Prior.
\newblock {\em Bayesian Analysis}, {\bf 8}(1), 111--132.

\bibitem[\protect\citename{Datta \& Ghosh, }2015]{datta2015search}
Datta, Jyotishka, \& Ghosh, Jayanta~K. 2015.
\newblock In Search of Optimal Objective Priors for Model Selection and
  Estimation.
\newblock {\em Current Trends in Bayesian Methodology with Applications},  225.

\bibitem[\protect\citename{Donoho, }2006]{donoho2006compressed}
Donoho, David~L. 2006.
\newblock Compressed Sensing.
\newblock {\em IEEE Transactions on information theory}, {\bf 52}(4),
  1289--1306.

\bibitem[\protect\citename{Donoho \& Johnstone, }1994]{donoho1994ideal}
Donoho, David~L., \& Johnstone, Iain~M. 1994.
\newblock Ideal spatial adaptation by wavelet shrinkage.
\newblock {\em Biometrika}, {\bf 81}(3), 425--455.

\bibitem[\protect\citename{Donoho \& Johnstone, }1995]{donoho1995adapting}
Donoho, David~L, \& Johnstone, Iain~M. 1995.
\newblock Adapting to unknown smoothness via wavelet shrinkage.
\newblock {\em Journal of the american statistical association}, {\bf 90}(432),
  1200--1224.

\bibitem[\protect\citename{Donoho {\em et~al.}, }1992]{donoho1992maximum}
Donoho, David~L, Johnstone, Iain~M, Hoch, Jeffrey~C, \& Stern, Alan~S. 1992.
\newblock Maximum Entropy and the Nearly Black Object.
\newblock {\em Journal of the Royal Statistical Society. Series B}, {\bf 54},
  41--81.

\bibitem[\protect\citename{Efron, }2008]{efron2008microarrays}
Efron, Bradley. 2008.
\newblock Microarrays, Empirical {{Bayes}} and the Two-Groups Model.
\newblock {\em Statistical Science}, {\bf 23}(1), 1--22.

\bibitem[\protect\citename{Efron, }2010]{efron2010large}
Efron, Bradley. 2010.
\newblock {\em Large-Scale Inference: Empirical {{Bayes}} Methods for
  Estimation, Testing, and Prediction}.
\newblock  Vol. 1.
\newblock {Cambridge University Press}.

\bibitem[\protect\citename{Efron {\em et~al.}, }2004]{efron_least_2004}
Efron, Bradley, Hastie, Trevor, Johnstone, Iain, \& Tibshirani, Robert. 2004.
\newblock Least Angle Regression.
\newblock {\em The Annals of statistics}, {\bf 32}(2), 407--499.

\bibitem[\protect\citename{Fan \& Li, }2001]{fan2001variable}
Fan, Jianqing, \& Li, Runze. 2001.
\newblock Variable Selection via Nonconcave Penalized Likelihood and Its Oracle
  Properties.
\newblock {\em Journal of the American statistical Association}, {\bf 96}(456),
  1348--1360.

\bibitem[\protect\citename{Faulkner \& Minin, }2015]{faulkner_bayesian_2015}
Faulkner, James~R., \& Minin, Vladimir~N. 2015.
\newblock Bayesian trend filtering: adaptive temporal smoothing with shrinkage
  priors.
\newblock Dec.

\bibitem[\protect\citename{Friedman {\em et~al.},
  }2007]{friedman_pathwise_2007}
Friedman, Jerome, Hastie, Trevor, H{\"o}fling, Holger, Tibshirani, Robert, \&
  {others}. 2007.
\newblock Pathwise Coordinate Optimization.
\newblock {\em The Annals of Applied Statistics}, {\bf 1}(2), 302--332.

\bibitem[\protect\citename{Friedman {\em et~al.}, }2008]{friedman2008sparse}
Friedman, Jerome, Hastie, Trevor, \& Tibshirani, Robert. 2008.
\newblock Sparse Inverse Covariance Estimation with the Graphical Lasso.
\newblock {\em Biostatistics}, {\bf 9}(3), 432--441.

\bibitem[\protect\citename{Friedman {\em et~al.},
  }2010]{friedman2010regularization}
Friedman, Jerome, Hastie, Trevor, \& Tibshirani, Rob. 2010.
\newblock Regularization Paths for Generalized Linear Models via Coordinate
  Descent.
\newblock {\em Journal of Statistical Software}, {\bf 33}(1), 1--22.

\bibitem[\protect\citename{Gelman {\em et~al.}, }2014]{gelman2014understanding}
Gelman, Andrew, Hwang, Jessica, \& Vehtari, Aki. 2014.
\newblock Understanding predictive information criteria for Bayesian models.
\newblock {\em Statistics and computing}, {\bf 24}(6), 997--1016.

\bibitem[\protect\citename{George, }2000]{george2000variable}
George, Edward~I. 2000.
\newblock The Variable Selection Problem.
\newblock {\em Journal of the American Statistical Association}, {\bf 95}(452),
  1304--1308.

\bibitem[\protect\citename{George \& Foster, }2000]{George0000}
George, EdwardI., \& Foster, Dean~P. 2000.
\newblock Calibration and Empirical {{Bayes}} Variable Selection.
\newblock {\em Biometrika}, {\bf 87}(4), 731--747.

\bibitem[\protect\citename{Ghosal {\em et~al.}, }2000]{ghosal2000}
Ghosal, Subhashis, Ghosh, Jayanta~K., \& {van der Vaart}, Aad~W. 2000.
\newblock Convergence Rates of Posterior Distributions.
\newblock {\em The Annals of Statistics}, {\bf 28}(2), 500--531.

\bibitem[\protect\citename{Ghosh {\em et~al.}, }2016]{ghosh2016testing}
Ghosh, Prasenjit, Tang, Xueying, Ghosh, Malay, \& Chakrabarti, Arijit. 2016.
\newblock Asymptotic properties of {B}ayes risk of a general class of shrinkage
  priors in multiple hypothesis testing under sparsity.
\newblock {\em Bayesian Analysis}, {\bf 11}(3), 753--796.

\bibitem[\protect\citename{Ghosh {\em et~al.}, }2017]{ghosh2016asymptotic}
Ghosh, Prasenjit, Chakrabarti, Arijit, {\em et~al.} 2017.
\newblock Asymptotic optimality of one-group shrinkage priors in sparse
  high-dimensional problems.
\newblock {\em Bayesian Analysis}, {\bf 12}(4), 1133--1161.

\bibitem[\protect\citename{Gordy, }1998]{gordy1998computationally}
Gordy, Michael~B. 1998.
\newblock Computationally convenient distributional assumptions for
  common-value auctions.
\newblock {\em Computational Economics}, {\bf 12}(1), 61--78.

\bibitem[\protect\citename{Gramacy {\em et~al.}, }2010]{gramacy2010shrinkage}
Gramacy, Robert~B, Pantaleo, Ester, \& {others}. 2010.
\newblock Shrinkage Regression for Multivariate Inference with Missing Data,
  and an Application to Portfolio Balancing.
\newblock {\em Bayesian Analysis}, {\bf 5}(2), 237--262.

\bibitem[\protect\citename{Griffin \& Brown, }2010]{griffin2005alternative}
Griffin, Jim~E, \& Brown, Philip~J. 2010.
\newblock Inference with Normal-Gamma Prior Distributions in Regression
  Problems.
\newblock {\em Bayesian Analysis}, {\bf 5}(1), 171--188.

\bibitem[\protect\citename{Hahn \& Lopes, }2014]{hahn_shrinkage_2014}
Hahn, P.~Richard, \& Lopes, Hedibert. 2014.
\newblock Shrinkage priors for linear instrumental variable models with many
  instruments.
\newblock {\em arXiv preprint arXiv:1408.0462}, Aug.

\bibitem[\protect\citename{Hahn {\em et~al.}, }2016]{hahn_elliptical_2016}
Hahn, P.~Richard, He, Jingyu, \& Lopes, Hedibert. 2016.
\newblock {\em Elliptical Slice Sampling for {{Bayesian}} Shrinkage Regression
  with Applications to Causal Inference}.
\newblock Tech. rept.

\bibitem[\protect\citename{Hahn {\em et~al.}, }2017]{hahn2017efficient}
Hahn, P~Richard, He, Jingyu, \& Lopes, Hedibert~F. 2017.
\newblock Efficient sampling for {G}aussian linear regression with arbitrary
  priors.

\bibitem[\protect\citename{Hans, }2011]{hans2011elastic}
Hans, Chris. 2011.
\newblock Elastic Net Regression Modeling with the Orthant Normal Prior.
\newblock {\em Journal of the American Statistical Association}, {\bf
  106}(496), 1383--1393.

\bibitem[\protect\citename{Hastie {\em et~al.}, }2009]{hastie_elements_2009}
Hastie, Trevor, Tibshirani, Robert, Friedman, Jerome, Hastie, T, Friedman, J,
  \& Tibshirani, R. 2009.
\newblock {\em The Elements of Statistical Learning}.
\newblock  Vol. 2.
\newblock {Springer}.

\bibitem[\protect\citename{Hastie {\em et~al.}, }2015]{hastie2015statistical}
Hastie, Trevor, Tibshirani, Robert, \& Wainwright, Martin. 2015.
\newblock {\em Statistical Learning with Sparsity}.
\newblock {CRC press}.

\bibitem[\protect\citename{Hoerl \& Kennard, }1970]{hoerl70}
Hoerl, Arthur~E., \& Kennard, Robert~W. 1970.
\newblock Ridge {{Regression}}: {{Biased Estimation}} for {{Nonorthogonal
  Problems}}.
\newblock {\em Technometrics}, {\bf 12}(1), 55--67.

\bibitem[\protect\citename{Ingraham \& Marks, }2016]{ingraham_bayesian_2016}
Ingraham, John~B., \& Marks, Debora~S. 2016.
\newblock Bayesian {Sparsity} for {Intractable} {Distributions}.
\newblock {\em arXiv preprint arXiv:1602.03807}.

\bibitem[\protect\citename{Ishwaran \& Rao, }2005]{ishwaran2005spike}
Ishwaran, Hemant, \& Rao, J.~Sunil. 2005.
\newblock Spike and slab variable selection: frequentist and {B}ayesian
  strategies.
\newblock {\em The Annals of Statistics}, {\bf 33}(2), 730--773.

\bibitem[\protect\citename{James {\em et~al.}, }2013]{james2013introduction}
James, Gareth, Witten, Daniela, Hastie, Trevor, \& Tibshirani, Robert. 2013.
\newblock {\em An Introduction to Statistical Learning}.
\newblock  Vol. 6.
\newblock {Springer}.

\bibitem[\protect\citename{James \& Stein, }1961]{james_estimation_1961}
James, William, \& Stein, Charles. 1961.
\newblock Estimation with Quadratic Loss.
\newblock {\em Pages  361--379 of:} {\em Proceedings of the Fourth {{Berkeley}}
  Symposium on Mathematical Statistics and Probability},  vol. 1.

\bibitem[\protect\citename{Javanmard \& Montanari,
  }2014]{javanmard2014confidence}
Javanmard, Adel, \& Montanari, Andrea. 2014.
\newblock Confidence intervals and hypothesis testing for high-dimensional
  regression.
\newblock {\em The Journal of Machine Learning Research}, {\bf 15}(1),
  2869--2909.

\bibitem[\protect\citename{Jeffreys \& Swirles, }1972]{jeffreys1972methods}
Jeffreys, Harold, \& Swirles, Bertha. 1972.
\newblock {\em {Methods of Mathematical Physics}}. 3 edn.
\newblock Cambridge: Cambridge university press.

\bibitem[\protect\citename{Johndrow \& Orenstein, }2017]{james2017scalable}
Johndrow, James~E., \& Orenstein, Paulo. 2017.
\newblock Scalable {{MCMC}} for {{Bayes Shrinkage Priors}}.
\newblock {\em arXiv preprint arXiv:1705.00841}.

\bibitem[\protect\citename{Johnstone \& Silverman, }2004]{johnstone2004needles}
Johnstone, Iain~M, \& Silverman, Bernard~W. 2004.
\newblock Needles and Straw in Haystacks: {{Empirical Bayes}} Estimates of
  Possibly Sparse Sequences.
\newblock {\em Annals of Statistics}, {\bf 32}, 1594--1649.

\bibitem[\protect\citename{Jolliffe {\em et~al.}, }2003]{jolliffe2003modified}
Jolliffe, Ian~T, Trendafilov, Nickolay~T, \& Uddin, Mudassir. 2003.
\newblock A Modified Principal Component Technique Based on the {{LASSO}}.
\newblock {\em Journal of computational and Graphical Statistics}, {\bf 12}(3),
  531--547.

\bibitem[\protect\citename{Kowal {\em et~al.}, }2017]{kowal2017dynamic}
Kowal, Daniel~R, Matteson, David~S, \& Ruppert, David. 2017.
\newblock Dynamic Shrinkage Processes.
\newblock {\em arXiv preprint arXiv:1707.00763}.

\bibitem[\protect\citename{Li, }1989]{li1989honest}
Li, Ker-Chau. 1989.
\newblock Honest confidence regions for nonparametric regression.
\newblock {\em The Annals of Statistics},  1001--1008.

\bibitem[\protect\citename{Li {\em et~al.}, }2017]{li2017graphical}
Li, Yunfan, Craig, Bruce~A, \& Bhadra, Anindya. 2017.
\newblock The Graphical Horseshoe Estimator for Inverse Covariance Matrices.
\newblock {\em arXiv preprint arXiv:1707.06661}.

\bibitem[\protect\citename{Liu \& Yu, }2013]{liu2013asymptotic}
Liu, Hanzhong, \& Yu, Bin. 2013.
\newblock Asymptotic properties of Lasso+ mLS and Lasso+ Ridge in sparse
  high-dimensional linear regression.
\newblock {\em Electronic Journal of Statistics}, {\bf 7}, 3124--3169.

\bibitem[\protect\citename{Louizos {\em et~al.}, }2017]{louizos2017bayesian}
Louizos, Christos, Ullrich, Karen, \& Welling, Max. 2017.
\newblock Bayesian compression for deep learning.
\newblock {\em Pages  3290--3300 of:} {\em Advances in Neural Information
  Processing Systems}.

\bibitem[\protect\citename{Magnusson {\em et~al.}, }2016]{magnusson_dolda_2016}
Magnusson, MÃ¥ns, Jonsson, Leif, \& Villani, Mattias. 2016.
\newblock {DOLDA} - a regularized supervised topic model for high-dimensional
  multi-class regression.
\newblock {\em arXiv preprint arXiv:1602.00260}, Jan.

\bibitem[\protect\citename{Makalic \& Schmidt, }2016]{makalic2016high}
Makalic, Enes, \& Schmidt, Daniel~F. 2016.
\newblock High-{{Dimensional Bayesian Regularised Regression}} with the
  {{BayesReg Package}}.
\newblock {\em arXiv preprint arXiv:1611.06649}.

\bibitem[\protect\citename{Mazumder {\em et~al.}, }2010]{mazumder2010spectral}
Mazumder, Rahul, Hastie, Trevor, \& Tibshirani, Robert. 2010.
\newblock Spectral Regularization Algorithms for Learning Large Incomplete
  Matrices.
\newblock {\em Journal of machine learning research}, {\bf 11}(Aug),
  2287--2322.

\bibitem[\protect\citename{Mazumder {\em et~al.}, }2012]{mazumder2012}
Mazumder, Rahul, Friedman, Jerome~H, \& Hastie, Trevor. 2012.
\newblock {{SparseNet}}: {{Coordinate}} Descent with Nonconvex Penalties.
\newblock {\em Journal of the American Statistical Association}, {\bf 106},
  1125--1138.

\bibitem[\protect\citename{Mitchell \& Beauchamp, }1988]{mitchell88}
Mitchell, T.~J., \& Beauchamp, J.~J. 1988.
\newblock Bayesian {{Variable Selection}} in {{Linear Regression}}.
\newblock {\em Journal of the American Statistical Association}, {\bf 83}(404),
  1023--1032.

\bibitem[\protect\citename{Nalenz {\em et~al.}, }2018]{nalenz2017tree}
Nalenz, Malte, Villani, Mattias, {\em et~al.} 2018.
\newblock Tree ensembles with rule structured horseshoe regularization.
\newblock {\em The Annals of Applied Statistics}, {\bf 12}(4), 2379--2408.

\bibitem[\protect\citename{Nickl {\em et~al.}, }2013]{nickl2013confidence}
Nickl, Richard, van~de Geer, Sara, {\em et~al.} 2013.
\newblock Confidence sets in sparse regression.
\newblock {\em The Annals of Statistics}, {\bf 41}(6), 2852--2876.

\bibitem[\protect\citename{Peltola {\em et~al.},
  }2014]{peltola2014hierarchical}
Peltola, Tomi, Havulinna, Aki~S, Salomaa, Veikko, \& Vehtari, Aki. 2014.
\newblock Hierarchical {{Bayesian}} Survival Analysis and Projective Covariate
  Selection in Cardiovascular Event Risk Prediction.
\newblock {\em Pages  79--88 of:} {\em Proceedings of the {{Eleventh UAI
  Conference}} on {{Bayesian Modeling Applications Workshop}}-{{Volume}} 1218}.
\newblock {CEUR-WS. org}.

\bibitem[\protect\citename{Piironen \& Vehtari,
  }2015]{piironen_projection_2015}
Piironen, Juho, \& Vehtari, Aki. 2015.
\newblock Projection predictive variable selection using {Stan}+{R}.
\newblock {\em arXiv preprint arXiv:1508.02502}, Aug.

\bibitem[\protect\citename{Piironen \& Vehtari, }2017a]{piironen2016hyperprior}
Piironen, Juho, \& Vehtari, Aki. 2017a.
\newblock On the Hyperprior Choice for the Global Shrinkage Parameter in the
  Horseshoe Prior.
\newblock {\em Pages  905--913 of:} {\em Artificial Intelligence and
  Statistics}.

\bibitem[\protect\citename{Piironen \& Vehtari, }2017b]{piironen2017sparsity}
Piironen, Juho, \& Vehtari, Aki. 2017b.
\newblock Sparsity information and regularization in the horseshoe and other
  shrinkage priors.
\newblock {\em Electronic Journal of Statistics}, {\bf 11}(2), 5018--5051.

\bibitem[\protect\citename{Polson \& Scott, }2010a]{polson2010large}
Polson, Nicholas~G, \& Scott, James~G. 2010a.
\newblock Large-Scale Simultaneous Testing with Hypergeometric Inverted-Beta
  Priors.
\newblock {\em arXiv preprint arXiv:1010.5223}.

\bibitem[\protect\citename{Polson \& Scott, }2010b]{polson2010shrink}
Polson, Nicholas~G, \& Scott, James~G. 2010b.
\newblock Shrink Globally, Act Locally: Sparse {{Bayesian}} Regularization and
  Prediction.
\newblock {\em Bayesian Statistics}, {\bf 9}, 501--538.

\bibitem[\protect\citename{Polson \& Scott, }2012a]{polson2012local}
Polson, Nicholas~G, \& Scott, James~G. 2012a.
\newblock Local shrinkage rules, L{\'e}vy processes and regularized regression.
\newblock {\em Journal of the Royal Statistical Society: Series B (Statistical
  Methodology)}, {\bf 74}(2), 287--311.

\bibitem[\protect\citename{Polson \& Scott, }2012b]{polson2012half}
Polson, Nicholas~G, \& Scott, James~G. 2012b.
\newblock On the Half-{{Cauchy}} Prior for a Global Scale Parameter.
\newblock {\em Bayesian Analysis}, {\bf 7}(4), 887--902.

\bibitem[\protect\citename{Polson {\em et~al.}, }2015]{polson2015proximal}
Polson, Nicholas~G, Scott, James~G, \& Willard, Brandon~T. 2015.
\newblock Proximal Algorithms in Statistics and Machine Learning.
\newblock {\em Statistical Science}, {\bf 30}(4), 559--581.

\bibitem[\protect\citename{Ro{\v{c}}kov{\'a} \& George,
  }2016]{rovckova2016spike}
Ro{\v{c}}kov{\'a}, Veronika, \& George, Edward~I. 2016.
\newblock The Spike-and-Slab Lasso.
\newblock {\em Journal of the American Statistical Association}.

\bibitem[\protect\citename{Scott, }2010]{scott_parameter_2010}
Scott, James~G. 2010.
\newblock Parameter Expansion in Local-Shrinkage Models.
\newblock {\em arXiv preprint arXiv:1010.5265}.

\bibitem[\protect\citename{Stein, }1956]{stein_inadmissibility_1956}
Stein, Charles. 1956.
\newblock Inadmissibility of the Usual Estimator for the Mean of a Multivariate
  Normal Distribution.
\newblock {\em Pages  197--206 of:} {\em Proceedings of the {{Third Berkeley}}
  Symposium on Mathematical Statistics and Probability},  vol. 1.

\bibitem[\protect\citename{Stephens \& Balding, }2009]{stephens2009bayesian}
Stephens, Matthew, \& Balding, David~J. 2009.
\newblock Bayesian Statistical Methods for Genetic Association Studies.
\newblock {\em Nature Reviews Genetics}, {\bf 10}(10), 681--690.

\bibitem[\protect\citename{Sun \& Zhang, }2012]{sun2012scaled}
Sun, Tingni, \& Zhang, Cun-Hui. 2012.
\newblock Scaled Sparse Linear Regression.
\newblock {\em Biometrika}, {\bf 99}(4), 879--898.

\bibitem[\protect\citename{Tang {\em et~al.}, }2016]{tang2016bayesian}
Tang, Xueying, Xu, Xiaofan, Ghosh, Malay, \& Ghosh, Prasenjit. 2016.
\newblock Bayesian Variable Selection and Estimation Based on Global-Local
  Shrinkage Priors.
\newblock {\em Sankhya A},  1--32.

\bibitem[\protect\citename{Terenin {\em et~al.},
  }2018]{terenin_gpu-accelerated_2016}
Terenin, Alexander, Dong, Shawfeng, \& Draper, David. 2018.
\newblock {GPU-accelerated Gibbs sampling: a case study of the Horseshoe Probit
  model}.
\newblock {\em Statistics and Computing},  1--10.

\bibitem[\protect\citename{Tiao \& Tan, }1966]{tiao1965bayesian}
Tiao, George~C., \& Tan, W.~Y. 1966.
\newblock Bayesian analysis of random-effect models in the analysis of
  variance. {II}. {E}ffect of autocorrelated errors.
\newblock {\em Biometrika}, {\bf 53}, 477--495.

\bibitem[\protect\citename{Tibshirani, }1996]{tibshirani96}
Tibshirani, R. 1996.
\newblock Regression {{Shrinkage}} and {{Selection}} via the {{Lasso}}.
\newblock {\em Journal of the Royal Statistical Society. Series B}, {\bf 58},
  267--288.

\bibitem[\protect\citename{Tibshirani {\em et~al.},
  }2005]{tibshirani_sparsity_2005}
Tibshirani, Robert, Saunders, Michael~A., Rosset, Saharon, Zhu, Ji, \& Knight,
  Keith. 2005.
\newblock Sparsity and Smoothness via the Fused Lasso.
\newblock {\em Journal of the Royal Statistical Society. Series B: Statistical
  Methodology}, {\bf 67}(1), 91--108.

\bibitem[\protect\citename{Tibshirani, }2014]{tibshirani2014praise}
Tibshirani, Robert~J. 2014.
\newblock In Praise of Sparsity and Convexity.
\newblock {\em Past, Present, and Future of Statistical Science},  497--505.

\bibitem[\protect\citename{Tibshirani \& Taylor, }2011]{tibshirani2011solution}
Tibshirani, Ryan~J., \& Taylor, Jonathan. 2011.
\newblock The solution path of the generalized lasso.
\newblock {\em Ann. Statist.}, {\bf 39}(3), 1335--1371.

\bibitem[\protect\citename{Tibshirani {\em et~al.},
  }2011]{tibshirani2011nearly}
Tibshirani, Ryan~J, Hoefling, Holger, \& Tibshirani, Robert. 2011.
\newblock Nearly-Isotonic Regression.
\newblock {\em Technometrics}, {\bf 53}(1), 54--61.

\bibitem[\protect\citename{Tikhonov, }1963]{tikhonov1963solution}
Tikhonov, Andrey. 1963.
\newblock Solution of incorrectly formulated problems and the regularization
  method.
\newblock {\em Soviet Meth. Dokl.}, {\bf 4}, 1035--1038.

\bibitem[\protect\citename{Tseng, }2001]{tseng2001convergence}
Tseng, Paul. 2001.
\newblock Convergence of a Block Coordinate Descent Method for
  Nondifferentiable Minimization.
\newblock {\em Journal of optimization theory and applications}, {\bf 109}(3),
  475--494.

\bibitem[\protect\citename{Van~de Geer {\em et~al.},
  }2014]{van2014asymptotically}
Van~de Geer, Sara, B{\"u}hlmann, Peter, Ritov, Ya’acov, Dezeure, Ruben, {\em
  et~al.} 2014.
\newblock On asymptotically optimal confidence regions and tests for
  high-dimensional models.
\newblock {\em The Annals of Statistics}, {\bf 42}(3), 1166--1202.

\bibitem[\protect\citename{{van der Pas} {\em et~al.}, }2014]{van2014horseshoe}
{van der Pas}, SL, Kleijn, BJK, \& {van der Vaart}, AW. 2014.
\newblock The Horseshoe Estimator: {{Posterior}} Concentration around Nearly
  Black Vectors.
\newblock {\em Electronic Journal of Statistics}, {\bf 8}, 2585--2618.

\bibitem[\protect\citename{{van der Pas} {\em et~al.},
  }2016a]{van2015conditions}
{van der Pas}, St{\'e}phanie, Salomond, Jean-Bernard, \& Schmidt-Hieber,
  Johannes. 2016a.
\newblock Conditions for {{Posterior Contraction}} in the {{Sparse Normal Means
  Problem}}.
\newblock {\em Electronic Journal of Statistics}, {\bf 10}, 976--1000.

\bibitem[\protect\citename{{van der Pas} {\em et~al.},
  }2016b]{pas_horseshoe:_2016}
{van der Pas}, Stephanie, Scott, James, Chakraborty, Antik, \& Bhattacharya,
  Anirban. 2016b.
\newblock {\em horseshoe: Implementation of the Horseshoe Prior}.
\newblock R package version 0.1.0.

\bibitem[\protect\citename{{van der Pas} {\em et~al.}, }2016c]{van2016many}
{van der Pas}, St{\'e}phanie, Szab{\'o}, Botond, \& {van der Vaart}, Aad.
  2016c.
\newblock How Many Needles in the Haystack? {{Adaptive}} Inference and
  Uncertainty Quantification for the Horseshoe.
\newblock {\em arXiv:1607.01892}.

\bibitem[\protect\citename{van~der Pas {\em et~al.}, }2017]{van2017adaptive}
van~der Pas, St{\'e}phanie, Szab{\'o}, Botond, van~der Vaart, Aad, {\em et~al.}
  2017.
\newblock Adaptive posterior contraction rates for the horseshoe.
\newblock {\em Electronic Journal of Statistics}, {\bf 11}(2), 3196--3225.

\bibitem[\protect\citename{Wang \& Pillai, }2013]{wang2013class}
Wang, Hao, \& Pillai, Natesh~S. 2013.
\newblock On a Class of Shrinkage Priors for Covariance Matrix Estimation.
\newblock {\em Journal of Computational and Graphical Statistics}, {\bf 22}(3),
  689--707.

\bibitem[\protect\citename{Wei, }2017]{wei2017bayesian}
Wei, Ran. 2017.
\newblock {\em Bayesian Variable Selection Using Continuous Shrinkage Priors
  for Nonparametric Models and Non-Gaussian Data.}
\newblock Ph.D. thesis, North Carolina State University.

\bibitem[\protect\citename{Witten {\em et~al.}, }2009]{witten2009penalized}
Witten, Daniela~M., Tibshirani, Robert, \& Hastie, Trevor. 2009.
\newblock A penalized matrix decomposition, with applications to sparse
  principal components and canonical correlation analysis.
\newblock {\em Biostatistics}, {\bf 10}(3), 515--534.

\bibitem[\protect\citename{Yuan \& Lin, }2006]{yuan2006model}
Yuan, Ming, \& Lin, Yi. 2006.
\newblock Model Selection and Estimation in Regression with Grouped Variables.
\newblock {\em Journal of the Royal Statistical Society: Series B (Statistical
  Methodology)}, {\bf 68}(1), 49--67.

\bibitem[\protect\citename{Zhang, }2010]{zhang2010nearly}
Zhang, Cun-Hui. 2010.
\newblock Nearly Unbiased Variable Selection under Minimax Concave Penalty.
\newblock {\em The Annals of Statistics}, {\bf 38}(2), 894--942.

\bibitem[\protect\citename{Zhang \& Zhang, }2014]{zhang2014confidence}
Zhang, Cun-Hui, \& Zhang, Stephanie~S. 2014.
\newblock Confidence intervals for low dimensional parameters in high
  dimensional linear models.
\newblock {\em Journal of the Royal Statistical Society: Series B (Statistical
  Methodology)}, {\bf 76}(1), 217--242.

\bibitem[\protect\citename{Zhang {\em et~al.}, }2016]{zhang2016high}
Zhang, Yan, Reich, Brian~J, \& Bondell, Howard~D. 2016.
\newblock High {{Dimensional Linear Regression}} via the {{R2}}-{{D2 Shrinkage
  Prior}}.
\newblock {\em arXiv preprint arXiv:1609.00046}.

\bibitem[\protect\citename{Zhao \& Yu, }2006]{zhao2006model}
Zhao, Peng, \& Yu, Bin. 2006.
\newblock On model selection consistency of Lasso.
\newblock {\em Journal of Machine learning research}, {\bf 7}(Nov), 2541--2563.

\bibitem[\protect\citename{Zou, }2006]{zou2006adaptive}
Zou, Hui. 2006.
\newblock The Adaptive Lasso and Its Oracle Properties.
\newblock {\em Journal of the American statistical association}, {\bf
  101}(476), 1418--1429.

\bibitem[\protect\citename{Zou \& Hastie, }2005]{zou2005regularization}
Zou, Hui, \& Hastie, Trevor. 2005.
\newblock Regularization and Variable Selection via the Elastic Net.
\newblock {\em Journal of the Royal Statistical Society: Series B (Statistical
  Methodology)}, {\bf 67}(2), 301--320.

\end{thebibliography}

\end{document}